\documentclass[prd,aps,eqsecnum,floatfix,nofootinbib,preprint,tightenlines]{revtex4}

\usepackage{latexsym}
\usepackage{graphicx}
\usepackage{multirow}
\usepackage{dcolumn}
\usepackage{hyperref}

\def\bibi{\bibitem}

\newcolumntype{k}{D{.}{.}{2.8}}
\newcolumntype{g}{D{.}{.}{3.6}}
\newcolumntype{a}{D{.}{.}{2.7}}

\let\ced=\c                     

\let\inodot=\i

\def\a{\alpha}
\def\b{\beta}
\def\c{\chi}
\def\d{\delta}
\def\g{\gamma}

\def\i{\iota}

\def\k{\kappa}

\def\m{\mu}
\def\n{\nu}

\def\p{\pi}                     
\def\r{\rho}                    
\def\t{\tau}

\def\D{\Delta}

\def\L{\Lambda}

\def\P{\Pi}

\def\co{{\cal O}}

\def\cq{{\cal Q}}

\def\cbo{{\,\raise-.15ex\Sc [\,}}                       

\def\ddt#1{{\buildrel {\hbox{\LARGE .\kern-2pt.}} \over {#1}}}

\def\ie{\mbox{\it i.e.}}
\def\eg{\mbox{\it e.g.}}

\def\floatcaption#1#2{ \caption{ #2 \ [#1] \label{#1}} }
\def\floatcaption#1#2{ \caption{#2 \label{#1}} }

\def\bibi{\bibitem}    

\def\ttl#1{{\it #1}}
\def\ttl#1{}

\long\def\symbolfootnote[#1]#2{\begingroup%
\def\thefootnote{\fnsymbol{footnote}}\footnote[#1]{#2}\endgroup}

\long \def \blockcomment #1\endcomment{}

\def\eg{{\it e.g.}}

\def\seef{{\it cf.}}

\def\bu{\overline{u}}

\def\hw{\hat{w}}

\def\textit#1{{\it #1}\kern.1em }

\def\ansatz{{\it ansatz}}

\begin{document}

\rightline{LMU-ASC 46/11}
\vspace{0.5cm}
\begin{center}
\begin{boldmath}
{\large\bf A new determination of  $\alpha_s$ from hadronic $\tau$ decays}\\[0.5cm]
\end{boldmath}
\vspace{3ex}
{ Diogo~Boito,$^a$ Oscar~Cat\`a,$^{b,c}$ Maarten~Golterman,$^d$ Matthias~Jamin,$^{e}$ Kim~Maltman,$^{f,g}$ James~Osborne,$^d$ Santiago~Peris$^d$%
\symbolfootnote[1]{Permanent address: Departament de F\'\inodot sica, Universitat Auton\`oma de Barcelona, E-08193 Bellaterra, Barcelona, Spain}
\\[0.3cm]
{\it
\null$^a$Departament de F\'\inodot sica and IFAE \\
Universitat Auton\`oma de Barcelona, E-08193 Bellaterra, Barcelona, Spain\\
\null$^b$Departament de F\'{\inodot}sica Te\`orica and IFIC\\
Universitat de Val\`encia-CSIC, E-46071 Val\`encia, Spain\\
\null$^c$Ludwig-Maximilians-Universit\"at M\"unchen, Fakult\"at f\"ur Physik\\
Arnold Sommerfeld Center for Theoretical Physics, D–80333 M\"unchen, Germany\\
\null$^d$Department of Physics and Astronomy\\
San Francisco State University, San Francisco, CA 94132, USA\\
\null$^e$Instituci\'o Catalana de Recerca i Estudis Avan\ced{c}ats (ICREA)\\
IFAE,  Universitat Auton\`oma de Barcelona\\ E-08193 Bellaterra, 
Barcelona, Spain\\
\null$^f$Department of Mathematics and Statistics\\ 
York University,  Toronto, ON Canada M3J~1P3\\
\null$^g$CSSM, University of Adelaide, Adelaide, SA~5005 Australia}}
\\[10mm]
{ABSTRACT}
\\[2mm]
\end{center}
\begin{quotation}
We present a new framework for the extraction of the strong coupling 
from hadronic $\tau$ decays through finite-energy sum rules. 
Our focus is on the small, but still significant non-perturbative effects 
that, in principle, affect both the central value and the systematic
error.  We employ a quantitative model in order to accommodate violations of
quark-hadron duality, and enforce a consistent treatment of the 
higher-dimensional contributions of the Operator Product Expansion to 
our sum rules. Using 1998 OPAL data for the non-strange isovector vector 
and axial-vector spectral functions, we find the $n_f=3$ values 
$\alpha_s(m_\tau^2)=0.307\pm 0.019$ in fixed-order perturbation theory, 
and $0.322\pm 0.026$ in contour-improved perturbation theory.
For comparison, the original OPAL analysis of the same data led to the 
values $0.324\pm 0.014$ (fixed-order) and $0.348\pm 0.021$ (contour-improved).
\end{quotation}

\vfill
\eject
\setcounter{footnote}{0}

\section{\label{intro} Introduction}
In the past few years there has been a renewed interest in the precision 
determination of $\a_s$ from non-strange hadronic $\t$ decays. One reason for
this interest is the recent calculation of the coefficient of the $O(\a_s^4)$
term in the perturbative contribution to the Adler function \cite{PT}. This
contribution dominates the ratio of the  hadronic $\tau$ decay width and the
electronic decay rate \cite{BNP},
\begin{equation}
\label{rtaudefn}
R_\tau = \Gamma [\tau^- \rightarrow \nu_\tau
\, {\rm hadrons}\, (\gamma)]/
\Gamma [\tau^- \rightarrow
\nu_\tau e^- {\bar \nu}_e (\gamma)]
\ .
\end{equation}
Another reason is the existence of a number of competing analysis methods 
which lead to results that are not, or only barely, consistent with 
one another. In fact, the error on $\a_s$ from $\t$ 
decays quoted in a recent (2009) review \cite{SB09} has gone 
$\it{up}$ since its 2006 version, for the simple reason that 
the result of Ref.~\cite{SB09} was obtained by averaging the
central values of all recent $\tau$ decay determinations of $\a_s$
and the error by considering the spread of these central values.
All determinations are
based on data from (primarily) the ALEPH (see Ref.~\cite{ALEPH98} for their
1998 analysis and Ref.~\cite{ALEPH} for their 2005 analysis) and (also) the
OPAL (see Ref.~\cite{OPAL})
collaborations; the differences in the results are on the theory side.

Clearly, this is an unsatisfactory situation. There are at least three
theoretical issues related to the discrepancies between the different
determinations. Number one is the long-standing question as to which
resummation scheme, fixed-order perturbation theory (FOPT) or contour-improved
perturbation theory (CIPT), is best used for evaluating the perturbative
contributions to $R_\t$. Many of the recent reanalyses have focussed on this
question \cite{PT,Davieretal08,BJ,Menke,CF,DM}. 
Much less attention has been devoted to the two other issues, both of which
concern non-perturbative contributions to $R_\t$. Because of the relatively
low value of the $\t$  mass, such contributions cannot be entirely neglected,
even if they are expected to be small. Issue number two is the question of
whether the Operator Product Expansion (OPE) contributions beyond 
perturbation theory have been consistently taken into account. Here 
we are aware of only one systematic investigation \cite{MY},
in which it was demonstrated that self-consistency problems existed
for a number of earlier analyses. Specifically,
it was shown that the OPE parameters obtained in those
analyses do not provide a good match between data and theory
when the upper limit $s_0$ on the hadronic invariant mass-squared, $s$, in the weighted integrals over the 
spectral functions (which enter
the finite-energy sum rules (FESRs) employed in these analyses)
is varied away from $m_\tau^2$.\footnote{Some of the earlier
analyses carried out this test for the FESR based on the 
kinematic weight (the weight yielding $R_\t$ when $s_0=m_\tau^2$), for 
which it works reasonably well.
Reference~\cite{MY} showed that these analyses unambiguously fail 
the tests for other doubly-pinched FESRs.}
Issue number three concerns potential violations of 
``quark-hadron duality'' not taken into account in previous 
FESR determinations of $\alpha_s$. In the case of FESRs, the assumption of 
quark-hadron duality amounts to presuming that all 
non-perturbative effects are accounted for by higher-dimensional 
terms in the OPE. To date, this assumption has not seen a 
systematic investigation. As already explained in Ref.~\cite{MY}, the 
issues of the OPE and possible violations of quark-hadron duality are
intricately connected: without a quantitative analysis of duality violations,
it turns out to be difficult to treat the OPE consistently without relying 
on ``external'' results.  For instance, in the analysis of Ref.~\cite{MY},
the dimension-4 term in the OPE had to be fixed using
a result for the gluon condensate from charmonium sum rules.

In this article, we aim to address this situation, focussing on the 
non-perturbative questions. We use a recently developed 
model for the duality-violating (DV) part of the $ud$-flavor vector ($V$) and 
axial-vector ($A$) spectral functions \cite{CGP}, which makes it possible to carry out 
a self-contained FESR analysis in which stability with
respect to varying $s_0$ is checked self-consistently without reliance on
external values for any of the OPE parameters.

Since no QCD-based theory of quark-hadron duality exists, we have to resort 
to a model. This means that our results will be based on the (testable)
assumption that this model gives a good description of the DV part of the 
spectral functions for values of $s$ from $s\to\infty$ down to
a minimum value, $s_{min}$, sufficiently low to lie in the range
$s\leq m_\tau^2$ kinematically accessible in hadronic $\tau$ decays. We
emphasize that the need to make such an assumption has always been a fundamental 
``shortcoming'' of the determination of $\a_s$ from $\t$ decays.   
Assuming quark-hadron duality {\it a priori}, and therefore neglecting
the effect of DVs altogether, also amounts to employing 
a -- probably worse -- model.  In other words, in order to 
investigate the systematics related to the assumption of quark-hadron 
duality, one cannot avoid the adoption of a model of the DV part of the 
spectral functions. In Ref.~\cite{CGP} it was found that the model
we intend to employ gives a reasonable description of the spectral 
functions in the region 
$1.1~\mbox{GeV}^2\le s\le m_\t^2$.\footnote{In the present, 
more detailed analysis, we 
will find that a significantly larger value of $s_{min}$ is preferred.}  
Moreover, the physics of our model is based on a picture of the hadron 
resonances which are experimentally seen in the spectral function. 
Resonances are not described by perturbation theory or the OPE, and thus 
should be part of any model aiming to describe violations of quark-hadron
duality.

In this work, we do not address the issue of the optimal choice of
resummation for the truncated perturbative series in a given FESR.
While this systematic, of course, forms a potentially important part of 
the final theory error on $\a_s$, we have no new elements to add to the 
discussion of this issue. Moreover, we believe that the non-perturbative 
part of the systematics should be understood first, in order to get a more 
reliable picture of the quantitative discrepancy between results based on FOPT and CIPT.
We will therefore carry out our whole analysis with 
both resummation schemes. 

To date, two experiments, ALEPH \cite{ALEPH98,ALEPH} and OPAL \cite{OPAL}, have 
made the non-strange $V$ and $A$ spectral functions from their 
$\t$-decay analyses publicly available. The 2005 analysis of ALEPH is more 
recent, and based on more statistics, and thus would be expected to have 
smaller experimental errors. Unfortunately, the 2005 ALEPH data cannot be used 
at present, because correlations due to unfolding have been inadvertently 
omitted in the original ALEPH analysis and hence from the publicly posted 
covariance matrices \cite{TAU2010}.\footnote{We thank members of the
ALEPH collaboration for private communications, in which the 
existence of this problem has been confirmed.}  Since the re-analysis of the ALEPH 
data has yet to be completed, we limit ourselves, in this article, to 
an analysis employing the OPAL data.

In order to normalize the various exclusive-mode components of the spectral 
functions, OPAL relied on the branching fractions available in 1998, as well 
as the then-current values of $V_{ud}$ and the electronic branching 
fraction $B_e$. All of these have been updated since then, and this makes 
it possible to at least partially update the OPAL inclusive
spectral distributions as well. Such an update would allow 
an updated, though still OPAL-based, determination of $\a_s$. 
Here, since our primary goal is to investigate the impact of
the novel features of our treatment of non-perturbative effects 
on the extracted results for $\alpha_s$, we choose {\it not} to 
perform this update, and instead work with the data in precisely the 
same form as used by OPAL \cite{OPAL}.  
We plan to devote a separate article to an adaptation of OPAL data to 
recent values of the exclusive branching fractions, $V_{ud}$ and $B_e$,
and an investigation of the effect of this adaptation on the value of 
$\a_s$ and other OPE parameters.

This article is organized as follows.
In Sec.~\ref{theory} we present a brief review of the application of 
FESRs to hadronic $\t$ decays, with emphasis on the issue of quark-hadron
duality. In Sec.~\ref{systematics} we are then able to provide a more thorough
discussion of the various systematic errors discussed already above.  
Preparing for a presentation of our results in Sec.~\ref{fits}, we describe 
the theory parametrization we will employ in more detail in 
Sec.~\ref{parametrization}, and discuss the issue of strong correlations
in the integrated data, and our resulting fitting strategies, in 
Sec.~\ref{data}. Apart from reporting on our fits in Secs.~\ref{w=1} and
\ref{multiple}, we consider also, in Sec.~\ref{V+A}, the $V+A$ channel sum 
(related to the non-strange part of $R_\t$) as well as the $V-A$ channel difference.
In the latter case, we demonstrate that
our fit results satisfy the 
Weinberg sum rules \cite{SW} as well as the
DGMLY sum rule for the $\pi$ electromagnetic mass difference \cite{EMpion}. 
Section~\ref{summary} contains a summary of our results, including 
a conversion of $\a_s$ to its value at the $Z$ mass; 
Sec.~\ref{conclusion} contains our conclusions.

\section{\label{theory} Theory summary}
Our analysis will involve the correlation functions
\begin{eqnarray}
\label{correl}
\P_{\m\n}(q)&=&i\int d^4x\,e^{iqx}\langle 0|T\left\{J_\m(x)J^\dagger_\n(0)\right\}|0\rangle\\
&=&\left(q_\m q_\n-q^2 g_{\m\n}\right)\P^{(1)}(q^2)+q_\m q_\n\P^{(0)}(q^2)\nonumber\\
&=&\left(q_\m q_\n-q^2 g_{\m\n}\right)\P^{(1+0)}(q^2)+q^2 g_{\m\n}\P^{(0)}(q^2)\ ,\nonumber
\end{eqnarray}
where $J_\m$ is one of the non-strange $V$ or
$A$ currents, $\bu\g_\m d$ or $\bu\g_\m\g_5 d$, and the superscripts
$(0)$ and $(1)$ label the spin. The decomposition in the third line employs the
combinations $\P^{(1+0)}(q^2)$ and $q^2\P^{(0)}(q^2)$, which 
are free of kinematic singularities.
Defining $s=q^2=\, -Q^2$ and the spectral functions
\begin{equation}
\label{spectral}
\r^{(1+0)}(s)=\frac{1}{\p}\;\mbox{Im}\,\P^{(1+0)}(s)\ ,
\end{equation}
Cauchy's theorem and the analytical properties of $\P^{(1+0)}(s)$,
applied to the contour in Fig.~\ref{cauchy-fig}, imply the FESR relation
\begin{eqnarray}
\label{cauchy}
I^{(w)}_{V/A}(s_0)\equiv\frac{1}{s_0}\int_0^{s_0}ds\,w(s)\,\r^{(1+0)}_{V/A}(s)
&=&-\frac{1}{2\p is_0}\oint_{|s|=s_0}
ds\,w(s)\,\P^{(1+0)}_{V/A}(s)\ ,
\end{eqnarray}
valid for any $s_0>0$ and any weight $w(s)$ analytic in the region of the 
contour \cite{shankar}. 
In the present work we will restrict ourselves to 
polynomial weights. Partial integration allows the right-hand side 
of Eq.~(\ref{cauchy}) to be recast in terms of the Adler function
\begin{equation}
\label{adler}
D(s)=-s\;\frac{d\P^{(1+0)}(s)}{ds}\ .
\end{equation}

\begin{figure}
\vspace*{4ex}
\begin{center}
\includegraphics*[width=6cm]{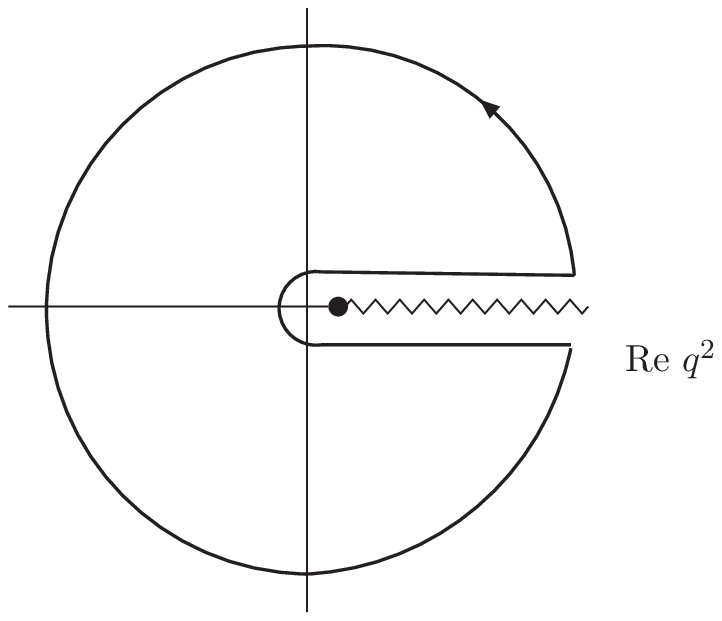}
\end{center}
\begin{quotation}
\floatcaption{cauchy-fig}%
{{\it Analytic structure of $\P^{(1+0)}(q^2)$ in the complex $s=q^2$ plane. The solid curve shows the contour used in Eq.~(\ref{cauchy}).}}
\end{quotation}
\vspace*{-4ex}
\end{figure}

The spectral functions $\rho^{(1+0)}_{V/A}(s)$ are measurable
in hadronic $\tau$ decays. Explicitly, for
Standard Model decays induced by the flavor $ud$ (isovector) currents, with
$w_T(s;s_0)=(1-s/s_0)^2(1+2s/s_0)$, 
$w_L(s;s_0)=2(s/s_0)(1-s/s_0)^2$, and
the scaled, non-strange $V$ and $A$ widths
\begin{equation}
\label{defR}
R_{V/A;ud}\equiv  {\frac{\Gamma [\tau^- \rightarrow \nu_\tau
\, {\rm hadrons}_{V/A;ud}\, (\gamma)]}
{\Gamma [\tau^- \rightarrow
\nu_\tau e^- {\bar \nu}_e (\gamma)]}}
\ ,
\end{equation}
one has \cite{tsai}
\begin{equation}
\label{RQCD}
R_{V/A;ud}=R_{V/A;ud}(s_0=m_\t^2)\ ,
\end{equation}
in which $R_{V/A;ud}(s_0)$ is defined by
\begin{equation}
\label{taukinspectral}
R_{V/A;ud}(s_0)= 12\pi^2\vert V_{ud}\vert^2 S_{EW}\,\frac{1}{s_0}
\int^{s_0}_0\, ds \,
\left[ w_T(s;s_0) \rho_{V/A}^{(1+0)}(s) - w_L(s;s_0) 
\rho_{V/A}^{(0)}(s) \right]\ ,
\end{equation}
where $S_{EW}$ is a short-distance electroweak correction.   Since,
in the Standard Model, 
\begin{eqnarray}
\label{scalar}
\rho_V^{(0)}(s)&=&O[(m_d-m_u)^2)]\ ,\\
\rho_A^{(0)}(s)&=&2f_\pi^2 \left(\d(s-m_{\pi^\pm}^2)
-\d(s)\right)+O[(m_d+m_u)^2]\ ,\nonumber
\end{eqnarray}
the differential distributions proportional to the expression in square brackets
in Eq.~(\ref{taukinspectral}) provide a direct
measure of $\rho_{V/A}^{(1+0)}(s)$, up to numerically negligible
$O\left(m_{u,d}^2\right)$ corrections.\footnote{The $\pi$ decay
constant, $f_\p=92.21(14)$~MeV \cite{PDG}, is presently known very accurately.
The central value of the
$\tau\rightarrow\pi\nu_\tau$ branching fraction, $B_\pi$, employed in the
1998 OPAL analysis corresponds to the somewhat
larger value $94.0$~MeV. In order to match exactly the OPAL treatment
of the $V$ and $A$ spectral functions we employ the latter value
in our analysis. Note that, since $B_\pi$ was obtained by the PDG
in a combined fit to the full set of $\tau$ basis modes, it would, in fact,
be inconsistent to change just $B_\pi$ without simultaneously changing
all other branching fractions.  We will revert
to the updated value $f_\pi =92.21(14)$~MeV in our later
analysis employing updated OPAL data.
}  The second delta function in Eq.~(\ref{scalar}), which comes from the kinematic
singularity present in $\P^{(0)}$, does not contribute to the integral
in Eq.~(\ref{taukinspectral}) as a result of the factor of $s$ in the accompanying weight
$w_L(s;s_0)$.

For sufficiently large $s_0$, and ignoring for a moment that the OPE is only
valid for large euclidean $Q^2$,
the right-hand side of Eq.~(\ref{cauchy}) can
be approximated using the OPE for $\Pi^{(1+0)}_{V/A}(s)$.
Experimental spectral data can then be used to fit the OPE, and extract 
parameters such as $\alpha_s$ \cite{BNP}.  
In what follows, we will denote the experimental version of the
spectral integral on the left-hand side of Eq.~(\ref{cauchy}) by $I_{V/A;ex}^{(w)}(s_0)$
(generically, $I_{ex}^{(w)}(s_0)$) and the theoretical
representation of the contour integral on the right-hand side by
        $I_{V/A;th}^{(w)}(s_0)$ (generically, $I_{th}^{(w)}(s_0)$). 

In the upper part of the energy region allowed by $\tau$-decay kinematics
$\left[ \Pi^{(1+0)}_{V/A}(s)\right]_{\rm OPE}$ is dominated by its
dimension $D=0$ contribution, \ie, the perturbative contribution in the chiral
limit.\footnote{Perturbative contributions proportional to powers of the quark
masses are included in $D\ge 2$ OPE terms.}   The perturbative expression
for the
Adler-function~(\ref{adler}), which is known to
order $\alpha_s^4$ \cite{PT}, can be written as\footnote{See, for instance,
Ref.~\cite{MJ}.}
\begin{equation}
\label{DPT}
[D(s)]^{D=0}_{\rm OPE}=\frac{1}{4\p^2}
\sum_{n=0}^\infty\;a_s^n(\m^2)\sum_{k=1}^{n+1}k\,c_{nk}
\left(\log\frac{-s}{\m^2}\right)^{k-1}\ ,
\end{equation}
with $a_s(\m^2)=\a_s(\m^2)/\p$. 
Since $D(s)$ is independent of $\m^2$, we can
choose  $\m^2=-s$, indicating that only
the coefficients $c_{n1}$ are independent; all other $c_{nk}$ can be expressed
in terms of the $c_{n1}$ through the renormalization group.  
In the $\overline{MS}$ scheme, $c_{01}=c_{11}=1$, $c_{21}=1.63982$,
$c_{31}=6.37101$ and $c_{41}=49.07570$ \cite{PT}.
We will use the guess $c_{51}=283$ of Ref.~\cite{BJ} for the next
coefficient, assigning an uncertainty of $\pm 283$ 
in order to estimate the error due to truncating 
perturbation theory (\seef\ Sec.~\ref{errors}).

The freedom to choose $\m^2$ in Eq.~(\ref{DPT}) is at the heart of the 
different prescriptions employed for evaluating the perturbative contribution
to the right-hand side of Eq.~(\ref{cauchy}):
in FOPT, $\mu^2=s_0$ is used
in Eq.~(\ref{DPT}), whereas in CIPT $\m^2=-s$ is employed 
inside the contour integral on the right-hand side of Eq.~(\ref{cauchy}) \cite{CIPT}.

Beyond perturbation theory, one may improve the approximation to the right-hand side of
Eq.~(\ref{cauchy}) by including higher-dimension
contributions to $\P^{(1+0)}_{\rm OPE}(s)$. Explicitly,
\begin{equation}
\label{OPE}
\P^{(1+0)}_{\rm OPE}(s)=\sum_{k=0}^\infty \frac{C_{2k}(s)}{(-s)^{k}}\ ,
\end{equation}
with the OPE coefficients $C_{2k}$ logarithmically dependent on $s$ through 
perturbative corrections. The term with $k=0$ corresponds
to the purely perturbative, mass-independent contributions,
represented by Eq.~(\ref{DPT}). 
The $C_{2k}$, with $k>1$, contain non-perturbative
$D=2k$ condensate contributions, and are, in principle, different 
for the $V$ and $A$ channels. 

We will neglect $C_2$, which is purely 
perturbative and quadratic in the light quark
masses.\footnote{Since in the present study we are 
only dealing with the light
up- and down-quark correlators, the $D=2$ mass-squared corrections
are tiny. Still, one version of our analysis code has implemented
all known $m^2$ corrections up to ${\co}(\a_s^3)$, and we
have verified that they are indeed negligible.}  
It has been suggested 
that a non-perturbative $D=2$ term should be added in order to 
account for the truncation of the perturbative series for the $D=0$ 
term \cite{NZ}.  We postpone an investigation of this issue to 
future work, and here set $C_2=0$.

Neglecting contributions of $O(m_{u,d}^4)$ or proportional to
$\langle \bar{u}u\rangle - \langle \bar{d}d\rangle$, both of which are numerically
very small,
the coefficient $C_4$
is a linear combination of the ``gluon condensate'' 
$\langle a_s G_{\m\n}G^{\m\n}\rangle$
(with $G_{\m\n}$ the gluon field strength), and the chiral condensates $m_i\langle
\overline{\psi}_i\psi_i\rangle$, $i=u,\ d,\ s$.  To leading order in $\a_s$ 
both contributions to $C_4$ are the same in the $V$ and $A$ channels.
Differences in the $D=4$ $V$ and $A$ light quark condensate contributions
enter beginning at $O(\alpha_s)$ or $O(m_{u,d}^4)$ \cite{NLOC4}. 

The coefficient $C_6$ is assumed to be dominated by four-quark condensates,
because the contribution from $\langle g^3f_{abc}G_\m^{a\n}G_\n^{b\k}G_\k^{c\m}
\rangle$ vanishes at leading order in $\a_s$ \cite{D6}, and the contributions from lower-dimensional operators are suppressed by powers of the quark mass.
Coefficient functions for the four-quark condensates were calculated to next-to-leading
order in Ref.~\cite{D64quark}.  At $D=8$ there is a proliferation of operators, and very
little detailed information is available.  As we will explain in Sec.~\ref{parametrization} below, we will not need to consider terms with $D>8$.

The OPE is valid when the euclidean distance $|x|$ in Eq.~(\ref{correl}) is small 
compared to $\L_{QCD}^{-1}$, or, equivalently, when euclidean
$Q^2$ is positive and large. However, both
perturbation theory and the OPE are expected to break down
near the positive real $q^2$ axis \cite{PQW}.
We may account for this additional non-perturbative effect by writing 
the right-hand side of Eq.~(\ref{cauchy}) as \cite{CGP}
\begin{equation}
\label{split}
-\frac{1}{2\p is_0}\oint_{|s|=s_0}ds\,w(s)\,\left(\P^{(1+0)}_{\rm OPE}(s)+\D(s)\right)\ ,
\end{equation}
with 
\begin{equation}
\label{DVdef}
\D(s)\equiv\P^{(1+0)}(s)-\P^{(1+0)}_{\rm OPE}(s)\ .
\end{equation}
The difference $\D(s)$ defines the duality violating 
contribution to $\Pi^{(1+0)}(s)$. 

All previous determinations of $\a_s$ from hadronic $\t$ decays have 
assumed, implicitly or explicitly, that integrated DV contributions 
are small enough to be neglected for the weights employed in the analysis.
While this assumption has sometimes been checked for self-consistency 
(see, \eg, Ref.~\cite{MY}), a comprehensive quantitative estimate of the impact of 
DVs on the precision with which $\a_s$ can be determined
has not been provided. One of the aims of the present work is to
provide a comprehensive analysis which takes DVs into account, and hence
remedies this shortcoming.  

As shown in Ref.~\cite{CGP}, if $\D(s)$ is assumed to decay
fast enough as $s\to\infty$, the right-hand side of the
FESR relation~(\ref{cauchy}) can be 
rewritten as
\begin{equation}
\label{sumrule}
I_{th}^{(w)}(s_0) = -\frac{1}{2\p is_0}\oint_{|s|=s_0}
ds\,w(s)\,\P^{(1+0)}_{\rm OPE}(s)+{\cal D}_w(s_0)\ ,
\end{equation}
with
\begin{equation}
\label{DV}
{\cal D}_w(s_0)=-\frac{1}{s_0}\int_{s_0}^\infty ds\,w(s)\,\frac{1}{\p}\,\mbox{Im}\,
\D(s)\ .
\end{equation}
The imaginary parts
$\frac{1}{\p}\,\mbox{Im}\,\D_{V/A}(s)$ can be interpreted as 
the DV parts, $\rho_{V/A}^{\rm DV}(s)$, of the $V/A$ spectral functions.
Following Ref.~\cite{CGP},
we will parametrize $\rho_{V/A}^{\rm DV}(s)$ as
\begin{equation}
\label{ansatz}
\rho_{V/A}^{\rm DV}(s)=
\k_{V/A}\,e^{-\g_{V/A}s}\sin{(\a_{V/A}+\b_{V/A}s)}\ .
\end{equation}
This introduces, in addition to $\a_s$ and the $D\ge 4$ OPE condensates, four 
new parameters in each channel. 
The OPE, supplemented by the \ansatz~(\ref{ansatz}),
will be assumed to hold for $s\ge s_{min}$, where $s_{min}$ will have 
to be inferred from fits to the experimental data. The extended analysis,
including DVs, will of course only be possible if $s_{min}$ lies
significantly below $m_\tau^2$.

The \ansatz~(\ref{ansatz}) is modeled on the asymptotic behavior for large $s$
of a semi-realistic
model for the QCD spectrum in a given channel. This model was developed in 
Ref.~\cite{CGPmodel}, based on earlier ideas described in Ref.~\cite{russians}.
It incorporates a combination of large-$N_c$ insights (narrow resonances 
with widths increasing with mass) as well as the Regge picture for the 
spacing between resonances. These ingredients lead quite naturally to
the exponential decay in Eq.~(\ref{ansatz}) with the decay
parameter $\g\sim 1/N_c$,
as well as the oscillatory behavior represented by the sine function, both
with arguments (approximately) linear in $s$.
We favor this model over other attempts to 
model DVs because of its natural connection to the resonance structure
of the spectral distributions, something that is not evident in other
models (such as those based on instantons). 
For detailed discussions of the model, see Refs.~\cite{CGPmodel,CGP05,MJ11}.

\section{\label{systematics} Systematic errors}
There are three sources of systematic error affecting, to various extents,
existing FESR determinations of $\a_s$. Since the investigation of two of these
sources is the central aim of this work, we briefly describe each of the three
sources here, before embarking on the details of our analysis.
\begin{itemize}
\item[1.]
There are (at least) two ways of partially resumming the perturbative
contribution to $I^{(w)}_{th}(s_0)$, CIPT and FOPT (\seef\ Sec.~\ref{theory}). 
The relative merits of the two methods have been the subject of a number
of investigations \cite{PT,Davieretal08,BJ,Menke,CF,DM}. While no particular
preference is given to either scheme in Refs.~\cite{PT,DM}, CIPT is favored
in \cite{Davieretal08,Menke}, whereas Refs.~\cite{BJ,CF} give arguments in
favor of FOPT, in the latter work through a new CI expansion in a conformally
mapped coupling.

We will not attempt to resolve the associated systematic uncertainty in this
work, but instead report on the results of our fits using both CIPT and FOPT.
In fact, it is interesting to see what discrepancy remains between CIPT and FOPT
after other systematic errors, described below, have been properly taken into 
account.
\item[2.]
With the exception of Ref.~\cite{MY}, the OPE has not been treated consistently
in previous extractions of $\a_s$ from $\t$ decays, in the sense we
now explain. Consider a term of order $1/s^k$ in the OPE of 
Eq.~(\ref{OPE}). The dominant term in the expansion in $\a_s$ of the corresponding 
coefficient $C_D$, $D=2k$, is a constant of order one 
(times the relevant condensate).
In the sum rule~(\ref{cauchy}), this term in the OPE is picked out 
by the term of degree $k-1$ in the weight $w(s)$. Other terms in $w(s)$ 
will also pick up contributions from $C_D(s)$, because of the logarithmic 
dependence of $C_D(s)$ on $s$, but such contributions will not be dominant 
as they are suppressed by at least one extra power of $\a_s$. Thus, if weight 
functions up to degree $n$ are used in the fits, it
follows that terms up to at least order $k=n+1$ must be kept in the OPE
in order to retain all potentially relevant
contributions not suppressed by at least one extra power of $\a_s$.

In the conventional analysis of Refs.~\cite{ALEPH,OPAL,Davieretal08,DP1992} this 
was not done: while weights up to degree $n=7$ in $s$ were employed (which 
would generally require keeping terms in the OPE up to $D=16$ ($k=8$)), 
only terms up to $D=8$ ($k=4$) were retained. As noted earlier,
it was found in Ref.~\cite{MY} that this is not self-consistent: with 
the parameter values found in those fits, the $s_0$ dependence
of the theory curves does not match that of the data for the majority 
of the weights employed, as well as for alternate degree 2 and 3 weights 
which explicitly test the $D\le 8$ parameters obtained from the original 
fits.\footnote{See also Ref.~\cite{CGPmodel}, where it was shown in a model
study that adding a $D=10$ term to the OPE in an analysis like that of 
Refs.~\cite{ALEPH,Davieretal08,OPAL} can make a significant difference.}
In this work, we will restrict ourselves to
weight functions of degree $n\le 3$, corresponding to keeping terms 
up to $D=8$ in the OPE. As we will explain below, this implies we 
will have to vary $s_0$ in the FESR~(\ref{cauchy}); it is not possible 
to restrict a consistent analysis to only $s_0=m_\t^2$ without 
additional uncontrollable assumptions.
\item[3.]
Typically, previous analyses neglected the presence of duality 
violations.\footnote{
Exceptions are Refs.~\cite{Davieretal08,Narison}; however, as is clear from 
the results of the present work, those investigations of DVs 
involved additional assumptions which we are able to avoid in the present
analysis.}  
While in some cases this assumption was checked for 
self-consistency \cite{MY}, such a check does not provide
a quantitative assessment of the impact of residual DV effects. It has 
long been known that FESRs with a simple weight like $w=1$ have sizable 
DVs, even at scales $\sim 2-3\ {\rm GeV}^2$, but that switching to 
weights ``pinched'' (having a zero) near $s=s_0$ significantly reduces 
this effect \cite{KM98DS99}. It has become a standard
assumption that using only weights which are at least doubly pinched will
suppress DVs sufficiently so as not to affect the value of, or error on, $\alpha_s$.
Clearly, in view of the quite small errors on $\a_s$ reported in the 
recent literature, this issue is in need of further investigation.

The use of weights which are at least doubly pinched,
\ie, which contain at least two powers of $s_0-s$, 
forces us to vary $s_0$ in Eq.~(\ref{cauchy}),
if the OPE is to be treated consistently in the sense described above. 
Suppose one wants to consider only $s_0=m_\t^2$, and fit $\a_s$ as 
well as $C_{4,\dots,D=2k}$, \ie, $k$ parameters.\footnote{Recall 
that we set $C_2=0$ in this article.} To fit $k$ parameters, one requires 
at least $k+1$ data points, and therefore at least $k+1$ linearly independent 
weights if one uses only $s_0=m_\t^2$. If all weights contain the 
factor $(s_0-s)^2$, the minimally required highest degree
will be $n=k+2$. This is inconsistent with our criterion of point 
2 above, which would require terms up to order $k+3$ to be kept in 
the OPE.  The only way out is to vary $s_0$, and/or to consider weights that
are less than doubly-pinched.
This makes the need to take DVs into account more urgent, because it is
certainly not justified to assume that integrated DVs are negligible 
for weights which are less than doubly pinched, over any sizable interval 
in $s_0$ below $m_\t^2$.
\end{itemize}

\section{\label{parametrization} Parametrization used in fits}
In this section, we describe in detail the parametrization of the theory that 
we will use in our analysis of the data.

\subsection{\label{which} Selection of moments}
We wish to consider terms in the OPE only up to $D=8$, for several reasons.
First, we expect that small contributions to typical OPE 
integrals associated with these condensates will be potentially sensitive 
to residual integrated DV contributions, making it possible
to check the impact of DVs on earlier determinations of the condensates. 
Second, it appears unlikely that we can reliably determine condensates 
with $D>8$ from existing data. Hence we focus on FESRs where such 
contributions will be strongly suppressed. Finally, little is known about 
the OPE, but it is almost certainly not a convergent
expansion. One would thus expect it to break down at 
some sufficiently high order, making it
prudent to limit ourselves to a relatively low maximum order.

{}From the arguments in Sec.~\ref{systematics}, this restricts us to weights with
degree $\le 3$, \ie, to at most four linearly independent weights. 
Since we will be fitting up to four OPE parameters ($\a_s$ as well as the 
$D=4,\ 6$ and $8$ condensates), in addition to the DV parameters in 
each channel, this already forces us to consider the sum rules~(\ref{cauchy}) 
with more than one value of $s_0$ for at least one weight. We will vary 
$s_0$ over the interval $[s_{min},m_\tau^2]$, and explore the stability 
of the fits as a function of $s_{min}$.  

In this work, we choose to consider the 
weights\footnote{We use hats in order to distinguish our set of weights from
a different set of weights considered in Ref.~\cite{MY}. }
\begin{eqnarray}
\hw_0(x)&=&1\ ,\label{weights}\\
\hw_2(x)&=&1-x^2\ ,\nonumber\\
\hw_3(x)&=&(1-x)^2(1+2x)=w_T(s;s_0)\ ,\nonumber\\
x&\equiv&s/s_0\ .\nonumber
\end{eqnarray}
A key point is that we explicitly incorporate DVs in our fits, and therefore 
need to use at least one weight sensitive not only to $\a_s$ and the OPE 
coefficients, but also to the DV parameters.  This stands in contrast to 
other work to date, where a desire to neglect DVs motivated the
use of (at least) doubly-pinched weights, which are known to suppress 
such contributions. Such doubly-pinched weights are, in fact, too 
insensitive to the DV parameters to allow for reliable fits of these 
parameters. In order to maximize our sensitivity to DVs and hence
improve our ability to fit DV parameters we include the unpinched
weight $\hat{w}_0$ in all our fits. The other weights have been chosen by 
requiring them to be of degree $\le 3$, to have no term linear in $s$, 
and to be singly pinched ($\hw_2$) or doubly pinched ($\hw_3$).   
Alternative sets of weights satisfying the same 
requirements are obtained by replacing either $\hw_2$ or $\hw_3$ 
with $1-x^3$. Our results with these alternative sets are completely 
consistent with those obtained from the set~(\ref{weights}).

For constant $C_D$, $I_{th}^{(\hw_2)}$ picks out the $D=6$ term, 
while $I_{th}^{(\hw_3)}$ picks out the  $D=6$ and $D=8$ terms. In practice, the
logarithmic dependence of $C_D$ on $s$ beyond leading order in $\alpha_s$
implies that all terms in the OPE contribute for all choices of $w(s)$.
However,  $C_6$ and $C_8$ will be primarily determined by  the $\hw_2$ and 
$\hw_3$ FESRs.
In the present work,
we will represent the $D\ge 4$ contributions using effective values $C_4$, $C_6$ and $C_8$ independent of $s$.  This implies that $C_4$ does not contribute to fits
involving any of the moments of Eq.~(\ref{weights}).
For the case of $C_4$, we have checked, in the case of fits to $I^{\hw_0}_{ex}(s_0)$,
that the numerical effect of this approximation is tiny, \seef\ Sec.~\ref{w=1}.

As we will show in Sec.~\ref{fits}, 
 fits to moments with weights $\hw_0$, $\hw_2$ and 
$\hw_3$ lead to stable and self-consistent results.   One might also consider including moments with weights containing
a linear term in $s$, such as the weight $w_1(x)=1-x$.
Such moments are sensitive
to the gluon condensate and would thus allow us to estimate its size,
although due care would have to be 
exercised with the
interpretation of such a result since this condensate mixes with the unit 
operator.\footnote{A similar observation holds
for some of the condensates in $C_6$ and $C_8$.} 
However,
renormalon-inspired model studies of higher orders in perturbation theory 
along the lines of Refs.~\cite{BJ,MJ} appear to indicate that perturbation 
theory, truncated at currently known orders, may be less 
well converged to the full resummed result for $I_{th}^{(w_1)}(s_0)$ than for 
$I_{th}^{(\hw_{0,2,3})}(s_0)$. This may be related to the observation that the
$D=4$ term, present in the OPE representation of $I_{th}^{(w_1)}(s_0)$, is 
affected by the leading infrared-renormalon ambiguity in the 
Borel resummation of the associated perturbative series. In contrast,
non-perturbative contributions to the  OPE representations of 
$I_{th}^{(\hw_{2,3})}(s_0)$
depend most significantly on the $C_D$ with $D>4$, and are affected primarily
only by subleading 
ambiguities, associated with more distant infrared renormalon poles.
This is one reason we restrict ourselves to the weights~(\ref{weights}) in this
article.

We have also studied weights with a term linear in $x$, such as $w_1(x)$,
added to our set of weights~(\ref{weights}),
but find that, with analyses of the type we present in Sec.~\ref{fits}, 
the uncertainty on the resulting determination of the gluon condensate
remains large compared to that of other determinations in the literature.
We intend to further investigate such fits, and in particular
the determination of the
gluon condensate, in a forthcoming article.   Here we just note that fits similar to those
presented in Sec.~\ref{fits} but also including the weight $w_1$ yield results consistent
with the fits presented in Secs~\ref{w=1} and~\ref{multiple}.
Finally, we observe that 
$\hat{w}_{0,2,3}(x)$ are a complete, linearly-independent set of weights 
of degree $\le 3$ without a term linear in $s$.

\subsection{\label{DVs} Duality violations}
We will parametrize the duality-violating part of $\r_V$ and $\r_A$
as in Eq.~(\ref{ansatz}).  This introduces four new parameters in each channel, 
forcing us to consider values of $s_0$ over an interval $[s_{min},m_\t^2]$.  
Since our \ansatz~(\ref{ansatz}) is assumed to hold only for sufficiently large
$s$, $s_{min}$ must be large enough 
to lie in the region of validity of this assumption, but low enough
to be kinematically accessible in hadronic $\tau$ decay.
Our \ansatz\ is therefore only practical if we also 
assume that such an $s_{min}$ exists. We are then interested in a value of  
$s_{min}$ that is small enough to maximize the data available 
for use in our fits, requiring at the same time that the DV \ansatz\ with that
choice of $s_{min}$ gives a good description of the data.

{\it A priori}, there is no reason for $s_{min}$ to be
equal in the $V$ and $A$ channels. We therefore present two types of 
analysis: one for the $V$ channel alone, and one for the combined 
$V$  and $A$ channels ($V\&A$). In the latter case, we will employ
an $s_{min}$ common to both channels. This is equivalent to assuming
an $s_{min}<m_\tau^2$ exists such that the asymptotic behavior has set in for 
both the $V$ and $A$ channels for all $s>s_{min}$. Since it seems unlikely 
for the asymptotic behavior in a given channel to set in below the 
lowest resonance in that channel, we expect to find
$s_{min}\sim m_{a_1}^2$ or higher for the combined $V\&A$ fits. 
In practice, we find an optimal choice $s_{min}\sim 1.4-1.5$~GeV$^2$. 

We have also considered fits to only the $A$ channel, but find that the data 
in that channel lead to a poor determination of the DV parameters, and 
thus also of $\a_s$. We believe this is due to the lower quality of the 
$A$-channel data, rather than the absence of a sufficiently low 
$A$-channel $s_{min}$, but it is impossible to decide this from the
data alone. We therefore do not discuss purely $A$-channel fits,
and restrict our analysis of this channel to combined fits 
involving the $V$ channel as well.

\section{\label{data} Correlations and fitting strategies}
Values of the left-hand side, $I_{ex}^{(w)}(s_0)$, of Eq.~(\ref{cauchy}) for nearby values of $s_0$
are very strongly correlated, and this has repercussions for the choice of 
fitting strategies. We describe the strong correlations in Sec.~\ref{corr}, and 
our strategies in Sec.~\ref{strategies} below.  As already explained in the
introduction, we limit ourselves to an analysis of the OPAL data \cite{OPAL}.

\subsection{\label{corr} Correlations and errors}
The data we will use are the OPAL compilation of the non-strange 
$V$ and $A$ spectral functions.\footnote{We would like to thank Sven Menke
for making the data files available to us.} These data 
appear in our fits through the weighted integrals, $I^{(w)}_{ex}(s_0)$, 
appearing on the left-hand sides of the FESRs~(\ref{cauchy}), for the 
various weight functions we consider. These integrals 
are, of course, represented numerically by sums over the appropriate
sets of experimental bins. Since OPAL's bin width is 0.032~GeV$^2$, 
varying $s_0$ between approximately 1.5~GeV$^2$
and 3.120~GeV$^2$ (OPAL's highest bin in the $V$ channel, 
and almost equal to $m_\t^2=3.157$~GeV$^2$) or 3.088~GeV$^2$ 
(OPAL's highest bin in the $A$ channel) provides about 50 data points 
for each integral.

The integrals $I^{(w)}_{ex}(s_0)$ are, however, highly correlated. 
For instance, if we consider $I^{\hw_0}_{ex}(s_0)$ on the interval 
$s_0\in[1.504,3.136]$~GeV$^2$,\footnote{These numbers are at the right edges of the bins centered at $s=1.488$~GeV$^2$
and $s=3.120$~GeV$^2$, respectively.} the 
corresponding correlation matrix has 
sub- or super-diagonal 
elements as large as $0.998$,
a largest eigenvalue 
$\sim 33$, and a smallest eigenvalue $\sim 0.00019$. Nonetheless,
as we will show in Sec.~\ref{w=1}, it turns out that
reliable, standard $\c^2$ fits to the data for $\hw_0$, using our 
parametrization of the  right-hand side of Eq.~(\ref{cauchy}), are possible.

The situation changes if we consider two or more weight functions 
simultaneously. Focussing on our primary fits, with moments constructed 
with the weight functions $\hw_{0,2,3}$, not only the correlations for 
each moment have to be taken into account, but also the cross-correlations 
between these different moments, because they are
not independent of each other.  In fact, if we consider the full correlation matrix for a
combination of moments, for a range of $s_0$ values, it turns out to
have zero eigenvalues at machine precision because of the strong 
cross-correlations. This means that standard $\c^2$ fits, employing
the full correlation matrix in constructing the function to be
minimized, are not possible in this case. This remains true if we 
``thin out'' the data, \ie, if we use fewer values of $s_0$ on a given 
interval, by a factor two to four. This puts standard $\c^2$ fits 
out of reach for simultaneous fits to multiple moments, forcing us to 
either use a different fitting strategy, or to drop such fits from 
consideration.

Because of this problem, we will perform fits to multiple moments 
using a different ``fit quality'' ${\cal Q}^2$.
${\cal Q}^2$ will be a positive-definite quadratic form in the differences 
between data and theory; for a description of some possible choices, see 
Sec.~\ref{strategies}. Any such ${\cal Q}^2$ can be minimized to give an estimate
of the fit parameters, as long as we have a reasonable way to estimate 
the parameter errors and covariances associated with the fit. In 
this article, we will estimate the parameter error matrix by propagating 
errors through a linear fluctuation analysis starting from the full data 
covariance matrix; for details, see App.~\ref{errorprop}. 

\subsection{\label{strategies} Fitting strategies}
In this subsection, we explain three different fitting strategies we have 
used, in view of 
the problem of strong correlations described in Sec.~\ref{corr}.  

\begin{itemize}
\item[1.]  The simplest fit we can perform is a standard $\c^2$ fit 
to a single moment. We will choose our single-moment fit to be the one 
with weight function $\hw_0(x)=1$, since it does not suppress contributions
from any part of the spectrum, and is sensitive enough to the DV part of our
fitting function to give reasonably good fits for the DV parameters. 
It should be noted here that our main goal is to minimize the fit error 
on $\a_s$. The only reason one cares about the ``nuisance'' parameters 
$\k_{V,A}$, $\g_{V,A}$, $\a_{V,A}$, and $\b_{V,A}$ is that they describe 
part of the physics, and as such have to be taken into account
in any fit. We emphasize again that, under the assumption that our DV 
\ansatz~(\ref{ansatz}) gives a good description of the DVs, 
there are no reasons to limit ourselves to pinched weights and, in fact,
strong arguments not to do so.

We have also considered fitting the spectral function directly, since 
it is maximally sensitive to DVs in the kinematically allowed
region. Such a fit can be cast in terms of an FESR 
obtained by choosing $w(s)=1$ and replacing $\r^{(1+0)}(s)$ with
its derivative on the left-hand side of Eq.~(\ref{cauchy}).
It turns out that it is not possible to determine $\a_s$ 
from such a fit, basically because the spectral function is much 
less sensitive to $\a_s$ than it is to the DV parameters.%
\footnote{In Ref.~\cite{CGP} fits to 
the spectral function were presented, but there $\a_s$ was kept fixed.
Note that contributions to $\rho_{V/A}$ from the higher $D$ terms in the OPE 
are negligibly small.} In contrast, pinched weights suppress the
contribution from DVs more than $\hw_0$ does. We find, in fact,
that fits with a single doubly-pinched weight are not stable if one
tries to fit both the OPE and (strongly suppressed) DV parameters.
In our experience, the most stable and precise results from a
single-moment fit are obtained using $\hw_0(x)=1$. Fits to singly-pinched 
weights also appear to work well: we have checked standard $\c^2$ fits 
with weights $\hw_2(x)$ or $1-x^3$ and find results in excellent agreement with
those reported in Tables~\ref{w0} and \ref{w023} below, 
though with somewhat larger errors on $\a_s$.
 
\item[2.] It is of course interesting to see whether simultaneous fits to 
multiple moments can be used to reduce errors, in particular the error on 
$\a_s$. However, in this case, we run into the problem of strong correlations 
described above in Sec.~\ref{corr}. The simplest solution is to omit 
correlations in constructing the fit quality ${\cal Q}^2$, and choose a 
${\cal Q}^2$ which is diagonal in the differences between data and theory.
Working with a set of $s_0$, $\{s_0^k\}$ in some fitting window,
and letting $\delta I_{ex}^{(w)}(s_0)$ be the error on the weighted
spectral integral $I_{ex}^{(w)}(s_0)$, obtained using the full
data covariance matrix, such a fit quality has the form
\begin{equation}
\label{diag}
{\cal Q}^2_{diag}=\sum_w\sum_{s_0^k} \left(\frac{I_{ex}^{(w)}(s_0^k)
-I_{th}^{(w)}(s_0^k;{\vec p})}{\delta I_{ex}^{(w)}(s_0^k)}\right)^2\ ,
\end{equation}
where we have made the dependence of the weighted theory integral
on the set of fit parameters $\vec{p}$ explicit and the outer
sum runs over the set of weights included in the analysis.
  
Often, such a fit is referred to as ``uncorrelated.'' Indeed, 
if ${\cal Q}^2$ would be interpreted as a $\c^2$, the standard 
$\c^2$ errors obtained from such a fit would miss the
effect of correlations and be (significantly) underestimated. 
However, we emphasize that we will not compute parameter errors in this 
way; instead we will propagate errors using the linear fluctuation analysis of
App.~\ref{errorprop}, thus taking into account all correlations 
explicitly.\footnote{We find indeed that this leads to much larger errors
than naive ``$\c^2$'' errors would suggest.}  

While we will not report on
fits to Eq.~(\ref{diag}), but instead rely on fits described under items 1 and 3,
we have carried out many such fits.  
They yield results fully consistent with those we do report, but always
lead to larger parameter errors.

\item[3.] A third type of fit we will consider is one in which 
${\cal Q}^2$ incorporates the correlation sub-matrix corresponding
to each individual moment employed in the fit, but not the 
cross-correlations between different moments. 
Full correlations are again to be included
via the linear fluctuation analysis described in App.~\ref{errorprop}. We choose
\begin{equation}
\label{blockcorr}
{\cal Q}^2_{block}=\sum_w\sum_{s_0^i,\, s_0^j}
\left(I_{ex}^{(w)}(s_0^i)-I_{th}^{(w)}(s_0^i;{\vec p})\right)
\left(C^{(w)}\right)^{-1}_{ij}
\left(I_{ex}^{(w)}(s_0^j)-I_{th}^{(w)}(s_0^j;{\vec p})\right)\ ,
\end{equation}
with $C_w$ the covariance matrix of the set of moments with fixed weight $w$
and $s_0$ running over the chosen fit window range.
The motivation for this form 
is that the cross-correlations between two moments arise mainly
because the weight functions used in multiple-moment fits are 
in practice close to being linearly dependent (even though, as 
a set of polynomials, of course they are not).
This dependency might be reinforced by the relatively large errors on the data
for values of $s$ toward $m_\t^2$, because it is primarily in this region 
that the weights $\hw_0$, $\hw_2$ and $\hw_3$ differ from each other.

\end{itemize}

A key observation is that it does not matter which fit quality 
${\cal Q}^2$ one chooses,\footnote{As long as ${\cal Q}^2$ is a 
positive-definite quadratic form in the differences
$I_{ex}^{(w)}(s_0^k)-I_{th}^{(w)}(s_0^k;\vec{p})$.}
as long as errors are propagated appropriately. Whatever the motivation 
for a particular choice, such a choice is useful if it turns out to 
allow a reliable fit, and to reduce errors on the fit parameters. 
We note that, of course, it is not possible to use the minimum value of 
diagonal or block diagonal ${\cal Q}^2$ of Eqs.~(\ref{diag}) and~(\ref{blockcorr}) obtained in such a fit in order to derive a confidence 
level; only the relative size of minimum values compared 
between different fits with the same choice of ${\cal Q}^2$ is meaningful.

\section{\label{fits} Fits}
In this section, we will present the results from our fits, using the 
parametrization of the theory explained in Sec.~\ref{parametrization} and 
employing the strategies of Sec.~\ref{strategies}. All fits are based on 
the original, unmodified OPAL data, including the OPAL normalization
for the $\pi$-pole contribution, which corresponds to a central value of $94.0$~MeV
for $f_\p$.

Section~\ref{w=1} contains our ``benchmark'' fit, which is a 
standard $\chi^2$ fit of $I_{ex}^{(\hw_0)}(s_0)$ for the $V$ channel. In this case,
the fit quality is the standard $\chi^2$ function, which
of course employs the full $I_{ex}^{(\hw_0)}(s_0)$ covariance matrix,
generated from the covariance matrix of the original OPAL data.
We will also consider a combined fit of the same moment to the $V$ and $A$ channels.

In Sec.~\ref{multiple} we consider simultaneous fits to $I_{ex}^{(\hw_0)}$, 
$I_{ex}^{(\hw_2)}$ and $I_{ex}^{(\hw_3)}$, again for both the pure 
$V$ channel and combined $V\& A$ channel cases.
As already discussed in Sec.~\ref{corr}, we find that standard $\chi^2$
fits are not possible. Our best results in the $V$ channel 
originate from minimizing the fit quality~(\ref{blockcorr}).  
Our main conclusion is that, presumably because of the strong correlations 
between different moments, these type of fits do not help reduce
the error on $\a_s$ significantly. They do, however, provide cross-checks, verifying that 
our \ansatz~(\ref{ansatz}) also describes moments other than just 
$I_{ex}^{(\hw_0)}$, including the non-strange component of
$R_\tau$, 
\begin{equation}
\label{RVpA}
R_{V+A;ud}(s_0)=R_{V;ud}(s_0)+R_{A;ud}(s_0)\ ,
\end{equation}
which is proportional to
$I_{ex,V+A}^{(\hw_3)}(s_0)$.  They also  give access
to the $V$- and $A$-channel 
OPE coefficients $C_{6,V}$, $C_{6,A}$, $C_{8,V}$ and $C_{8,A}$.

In Sec.~\ref{errors} we consider the additional errors originating from the 
truncation of perturbation theory, and in Sec.~\ref{V+A} we show that our 
fits both give a good description of $R_{V+A;ud}$, 
and satisfy, within errors, the Weinberg and DGMLY $V-A$ sum-rule
constraints.   

\begin{boldmath}
\subsection{\label{w=1} Fits with the weight $\hw_0(x)$}
\end{boldmath}
We begin with a standard $\chi^2$ fit to the $V$-channel
$w(s)=\hw_0(x)=1$ FESR. 
Fit results are presented in Table~\ref{w0},
which shows all fit parameters, as well as the number of degrees of freedom
(dof), and the $\c^2$ per degree of freedom. 
Errors are standard $\c^2$ errors; errors computed with Eq.~(\ref{parcorr}) are typically somewhat larger, but 
similar in size.
\begin{table}[t]
\begin{center}
\begin{tabular}{|c|c|c|a|k|g|g|g|}
\hline
$s_{min}$ & {\rm dof} &$\c^2$/{\rm dof} & \multicolumn{1}{c|}{$\alpha_s$} & \multicolumn{1}{c|}{$\kappa_V$} & \multicolumn{1}{c|}{$\gamma_V$} & \multicolumn{1}{c|}{$\alpha_V$} & \multicolumn{1}{c|}{$\beta_V$} \\
\hline
1.3 & 53 & 0.44 & 0.320(23) & 0.026(18) & 0.42(50) & 0.54(54) & 2.85(30) \\
1.4 & 50 & 0.35 & 0.311(19) & 0.019(13) & 0.23(44) & -0.29(64) & 3.27(33) \\
1.5 & 47 & 0.36 & 0.307(18) & 0.017(11) & 0.16(42) & -0.52(74) & 3.38(38) \\
1.6 & 44 & 0.38 & 0.308(20) & 0.018(15) & 0.22(51) & -0.47(82) & 3.36(41) \\
1.7 & 41 & 0.38 & 0.305(19) & 0.012(12) & 0.03(51) & -0.61(86) & 3.41(41) \\
\hline
\hline
1.3 & 53 & 0.45 & 0.332(37) & 0.037(27) & 0.64(53) & 0.57(58) & 2.80(33) \\
1.4 & 50 & 0.36 & 0.326(27) & 0.023(16) & 0.35(48) & -0.32(64) & 3.27(33) \\
1.5 & 47 & 0.37 & 0.322(25) & 0.020(13) & 0.25(44) & -0.57(73) & 3.39(38) \\
1.6 & 44 & 0.38 & 0.323(27) & 0.022(20) & 0.31(57) & -0.53(81) & 3.38(41) \\
1.7 & 41 & 0.39 & 0.320(25) & 0.014(13) & 0.08(53) & -0.68(85) & 3.43(40) \\
\hline
\end{tabular}
\end{center}
\begin{quotation}
\floatcaption{w0}{{\it Standard $\chi^2$ fits to 
Eq.~(\ref{cauchy}) with $w(s)=1$, V channel. 
FOPT results are shown above the double horizontal line, CIPT results below.
Errors are standard $\c^2$ errors;
$\g_V$ and $\b_V$ in {\rm GeV}$^{-2}$. }}
\end{quotation}
\vspace*{-4ex}
\end{table}%

\begin{figure}[t]
\centering
\includegraphics[width=2.9in]{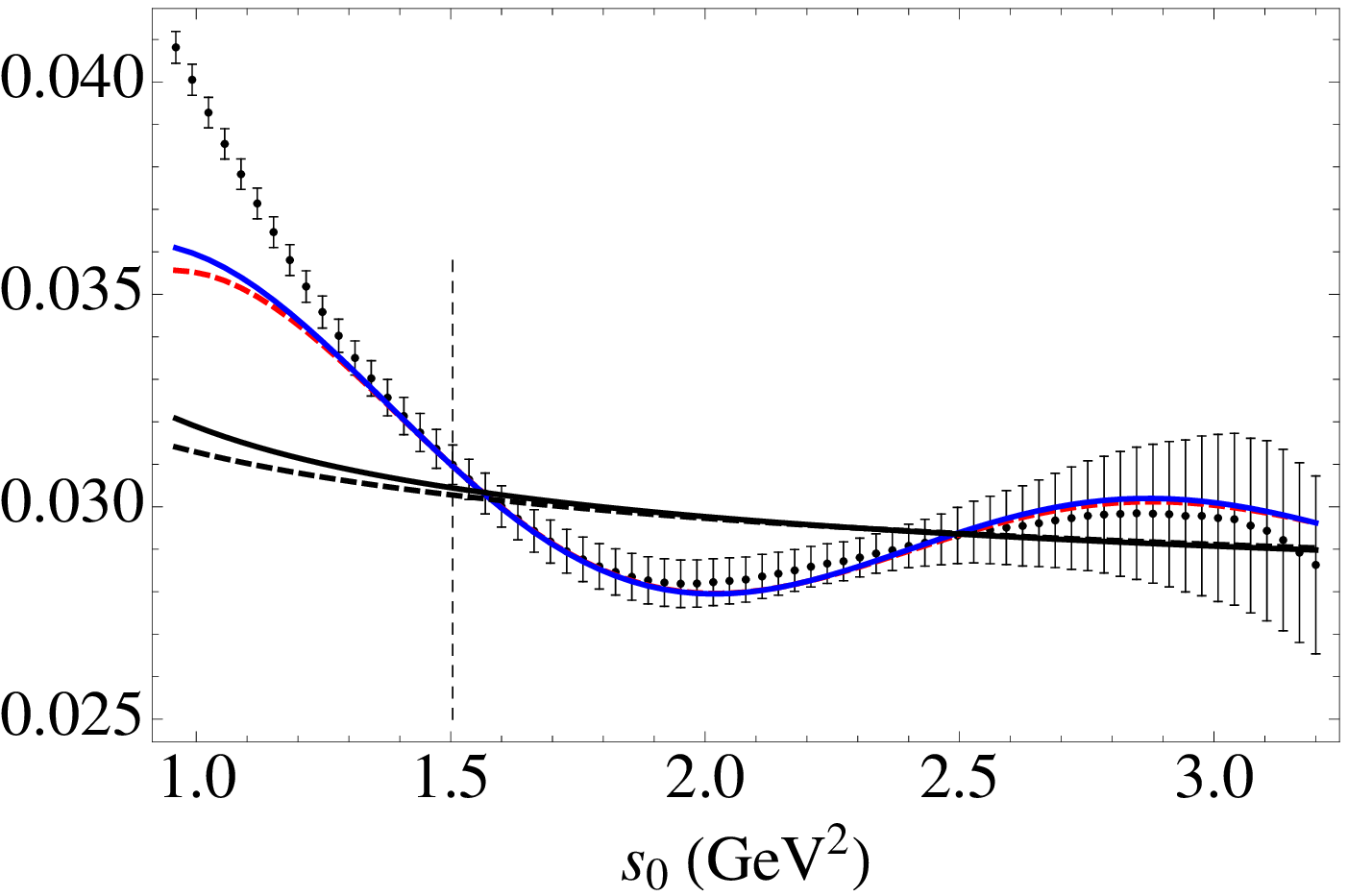}
\hspace{.1cm}
\includegraphics[width=2.9in]{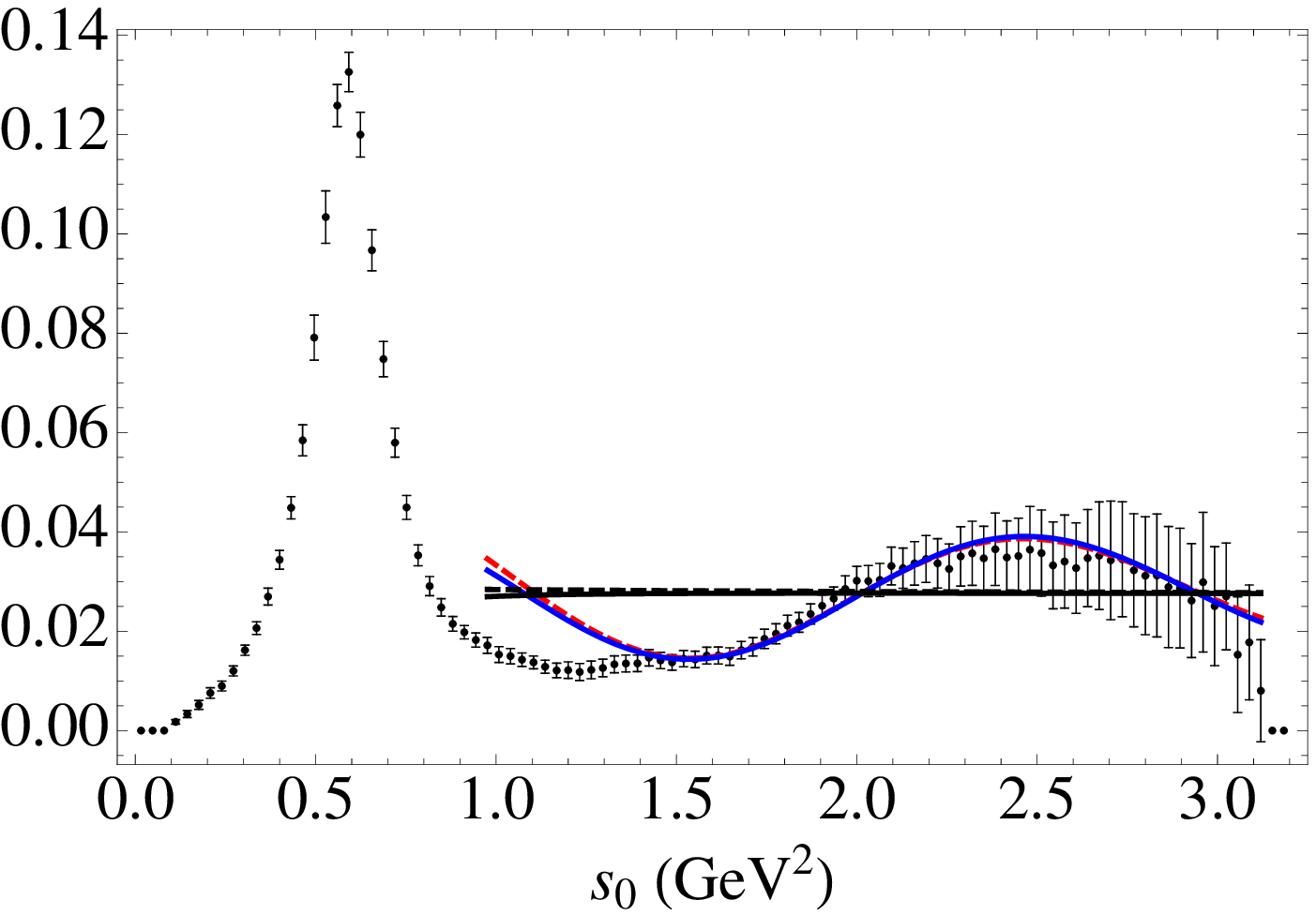}
\floatcaption{Vw0figure}{\it Left panel: 
comparison of $I^{(\hw_0)}_{ex}(s_0)$
       and $I^{(\hw_0)}_{th}(s_0)$ for the $s_{min}=1.5\ {\rm GeV}^2$
       V-channel fits of Table~\ref{w0}.
   Right panel: comparison
of the theoretical spectral function resulting
from this fit with the experimental results.  CIPT fits are shown in red (dashed) and 
FOPT in blue (solid).
The (much flatter) black curves display only the OPE parts of the
FOPT (solid) and CIPT (dashed) fit results.  The vertical dashed line
indicates the location of $s_{min}$.}
\vspace*{2ex}
\end{figure}

We observe 
that there is excellent stability for the results with $s_{min}=1.4$, 
$1.5$ and $1.6$~GeV$^2$, with the errors getting somewhat larger at 1.6~GeV$^2$. 
At values of $s_{min}\ge 1.7$~GeV$^2$, $\g_V$ becomes very small and tends
to go negative, which is clearly unphysical.  However, within errors such
fits are always consistent with those shown in the table.
We choose the results obtained at $s_{min}=1.5$~GeV$^2$ 
to fix our central values, and treat the spread of
the fit results for $s_{min}$ ranging from 1.4 to 1.6~GeV$^2$ as an error.
{}From this simple fit, we obtain for $\a_s$ at the $\t$ mass
the values
\begin{eqnarray}
\label{asw0}
\a_s(m_\t^2)&=&0.307\pm 0.018\pm 0.004\qquad\mbox{(FOPT)}\ ,\\
\a_s(m_\t^2)&=&0.322\pm 0.025\pm 0.004\qquad\mbox{(CIPT)}\ ,\nonumber
\end{eqnarray}
where the second error represents the variation with $s_{min}$ 
discussed above. We note that there is good stability over the
full range of $s_{min}$ values covered in Table~\ref{w0}.
Since our fitting function is non-linear, in general $\c^2$ errors are expected
to be asymmetric. We have therefore also computed asymmetric errors for all the fit parameters.
We find that the error on $\a_s$ is nearly symmetric, and that only errors
on $\k_V$ and $\g_V$ show a significant asymmetry.  For instance, we find,
at $s_{min}=1.5$~GeV$^2$, that, for FOPT,
\begin{eqnarray}
\label{kappa}
\a_s(m_\t^2)&=&0.307^{\,+0.018}_{\,-0.021}\ ,\\
\k_V&=&0.017^{\,+0.027}_{\,-0.007}\ ,\nonumber\\
\g_V&=&0.16^{\,+0.63}_{\,-0.34}\ \mbox{GeV}^{-2}\ ,\nonumber\\
\a_V&=&-0.52^{\,+0.71}_{\,-0.78}\ ,\nonumber\\
\b_V&=&3.38^{\,+0.40}_{\,-0.36}\ \mbox{GeV}^{-2}\ .\nonumber
\end{eqnarray}
This shows that omitting DVs from the fit, which is equivalent to setting
$\k_V=0$, would lead to a poor fit.  For CIPT the asymmetries show the same
pattern.

\begin{table}[t]
\begin{center}
\begin{tabular}{|c|ccccc|}
\hline
 & $\alpha_s$ & $\kappa_V$ & $\gamma_V$ & $\alpha_V$ & $\beta_V$ \\
\hline
$\alpha_s$ & 1 & -0.69 & -0.67 & 0.70 & -0.62 \\
$\kappa_V$ & -0.69 & 1 & 0.99 & -0.47 & 0.43 \\
$\gamma_V$ & -0.67 & 0.99 & 1 & -0.48 & 0.43 \\
$\alpha_V$ & 0.70 & -0.47 & -0.68 & 1 & -0.98 \\
$\beta_V$ & -0.62 & 0.43 & 0.43 & -0.98 & 1 \\
\hline
\end{tabular}
\end{center}
\begin{quotation}
\floatcaption{w0parcorr}{{\it Parameter correlation matrix for the FOPT fit 
with $s_{min}=1.5$~{\rm GeV}$^2$ shown in
Table~\ref{w0}.}}
\end{quotation}
\vspace*{-4ex}
\end{table}%

In Fig.~\ref{Vw0figure}, we show the quality of the match between the fitted
$I_{th}^{(\hat{w}_0)}(s_0)$ and $I_{ex}^{(\hat{w}_0)}(s_0)$, as well as 
of the match between the experimental spectral function and its theoretical
counterpart in which parameter values are obtained from the FESR fit  
for $s_{min}=1.5$~GeV$^2$.   The left panel shows the FESR fit itself, while
the right panel
shows the results of this comparison for the spectral function case. 
We emphasize that the spectral function was not part of the fit;
the agreement between the theoretical and experimental versions of $\rho_V(s)$
is an output.  
Agreement with data is good in the full fit window
$s_0\geq s_{min}= 1.5$~GeV$^2$. The black curves, in contrast, 
show the OPE parts of the theoretical curves, \ie, the curves 
obtained by removing the DV contributions from the blue and red
curves. It is clear that DVs are needed to give a good description 
of the data for $I_{ex}^{(\hw_0)}$ and the spectral function itself,
in agreement with our conclusion based on Eq.~(\ref{kappa}).
 
There are strong correlations between the parameters shown in 
Table~\ref{w0} and Eq.~(\ref{kappa}). Such strong correlations are present 
in all our fits, and are unavoidable, given the number of fit 
parameters.  We emphasize that this cannot be resolved by simply 
omitting the duality-violating part from the
theory -- one cannot ``improve'' fits by throwing out physics that is known to 
have an impact on those fits! In Table~\ref{w0parcorr} we show the full
parameter correlation matrix corresponding to the FOPT result
quoted in Eq.~(\ref{asw0}). The analogous matrix for the CIPT result looks very
similar.

We have also considered fits like those shown in Table~\ref{w0},
but including 
the contribution coming from
the logarithmic dependence of $C_4$ on $s$.  For the latter, we estimated
the quark-condensate contribution from the Gell-Mann--Oakes-Renner relation
\cite{GMOR},
and we took $\langle a_sG_{\m\n}G^{\m\n}\rangle=0.021$~GeV$^4$.  We find that 
corresponding changes in the numbers in Table~\ref{w0} are at most a tiny
fraction of the fitting errors.

We performed a similar type of fit to the combination of $V$ and $A$ channels.
If we use all possible $s_0$
values, we find that the standard $\c^2$ 
fit function (involving the very strongly correlated spectral integral
covariance matrix) is very flat, admitting not just ``physical'' solutions 
(consistent with those in Table~\ref{w0}), but also solutions 
that are clearly unphysical (with, for instance, values for $\a_s$ 
drifting down to unacceptably low values as a function of $s_{min}$). 
This happens for both CIPT and FOPT. Parameter errors
for these solutions can be very large, consistent with the flatness 
of the $\c^2$ landscape, and there is a strong sensitivity of central 
values to initial guesses for the parameter values.

However,
if we thin out the $s_0$ values
\ie, use only every $n$th value of $I^{\hw_0}_{ex}(s_0)$,
for some value of $n>1$, we find that fits  with $n=2$, $3$ or $4$ are much
more stable than the one with $n=1$ (no thinning) described above.  In Table~\ref{VAw0} we show our results for $n=3$, which is the choice leading to the most stable fits.\footnote{Negative values for $\g_{V,A}$ can appear because we use the analytic form of the
integral in Eq.~(\ref{DV}).  Note, however, that all values we find in the fits are consistent
with a positive value, as physically required.}
The problem disappears when we use fit quality~(\ref{diag}), computing
errors with Eq.~(\ref{parcorr}), but this method leads to significantly larger errors.

Choosing again the fit with $s_{min}=1.5$~GeV$^2$, we find for our combined
$V$ and $A$ channel fit the results
\begin{eqnarray}
\label{asw0VA}
\a_s(m_\t^2)&=&0.308\pm 0.016\pm 0.009\qquad\mbox{(FOPT)}\ ,\\
\a_s(m_\t^2)&=&0.325\pm 0.022\pm 0.011\qquad\mbox{(CIPT)}\ ,\nonumber
\end{eqnarray}
with the second error representing, as above, the variation 
over the neighboring $s_{min}$ values, $1.4$ and $1.6$ GeV$^2$. The results 
reported in Tables~\ref{w0} and \ref{VAw0} are 
in good agreement.   We note that the central values for $\g_V$
shown in Table~\ref{VAw0} are very small, compared with those
in Table~\ref{w0}, but given the errors there is no inconsistency.
In Fig.~\ref{VAw0figure} we show the quality 
of the fit in the panels on the left, for $V$ (top) and $A$ (bottom), for
$s_{min}=1.5$~GeV$^2$.  In the panels
on the right we show again the match between the experimental spectral
functions and their theoretical counterparts with parameter values obtained 
from the $s_{min}=1.5$~GeV$^2$ FESR fit.  As before, black curves show only the
OPE parts of the theoretical curves.
Again, we see that the data clearly confirm the presence of DVs, which 
are well described by our \ansatz.

\begin{table}[t]
\begin{center}
\begin{tabular}{|c|c|c|a|k|g|g|g|}
\hline
$s_{min}$ & {\rm dof} & $\c^2$/{\rm dof} & \multicolumn{1}{c|}{$\a_s$} & \multicolumn{1}{c|}{$\k_{V,A}$} & \multicolumn{1}{c|}{$\g_{V,A}$} & \multicolumn{1}{c|}{$\a_{V,A}$} & \multicolumn{1}{c|}{$\b_{V,A}$} \\
\hline
1.3 & 30 & 0.81 & 0.326(13) & 0.0186(88) & 0.18(35) & 0.35(46) & 2.95(27) \\
&&&& 0.094(51) & 1.15(35) & 0.21(79) & -3.42(45) \\
\hline
1.4 & 28 &  0.69 & 0.317(15) & 0.0140(68) & 0.01(33) & -0.31(58) & 3.29(31) \\
&&&& 0.085(39) & 1.06(30) & -0.5(1.1) & -3.06(61) \\
\hline
1.5 & 26 & 0.69 & 0.308(16) & 0.0134(69) & -0.01(34) & -0.67(70) & 3.46(36) \\
&&&& 0.110(71) & 1.15(35) & -1.1(1.1) & -2.73(59) \\
\hline
1.6 & 24 & 0.73 & 0.308(17) & 0.0150(98) & 0.06(41) & -0.64(74) & 3.45(38) \\
&&&& 0.15(13) & 1.29(45) & -1.2(1.2) & -2.67(65) \\
\hline
1.7 & 22 & 0.68 & 0.304(18) & 0.0131(99) & 0.00(42) & -0.80(77) & 3.51(38) \\
&&&& 1.0(2.0) & 2.14(84) & -2.3(2.1) & -2.1(1.1) \\
\hline
\hline
1.3 & 30 & 0.85 & 0.346(19) & 0.026(12) & 0.37(34) & 0.40(51) & 2.89(30) \\
&&&& 0.072(35) & 1.01(31) & 0.10(85) & -3.38(48) \\
\hline
1.4 & 28 & 0.72  & 0.336(21) & 0.0170(86) & 0.10(35) & -0.36(56) & 3.30(30) \\
&&&& 0.075(33) & 0.99(28) & -0.5(1.1) & -3.04(61) \\
\hline
1.5 & 26 & 0.71 & 0.325(22) & 0.0152(80) & 0.05(35) & -0.73(67) & 3.48(35) \\
&&&& 0.101(66) & 1.11(35) & -1.2(1.0) & -2.74(58) \\
\hline
1.6 & 24 & 0.75 & 0.324(23) & 0.017(12) & 0.11(44) & -0.72(71) & 3.47(37) \\
&&&& 0.14(12) & 1.25(45) & -1.3(1.1) & -2.66(62) \\
\hline
1.7 & 22 & 0.70 & 0.318(23) & 0.014(11) & 0.03(44) & -0.88(75) & 3.54(37) \\
&&&& 1.0(1.9) & 2.11(86) & -2.3(2.1) & -2.1(1.1) \\
\hline
\end{tabular}
\end{center}
\begin{quotation}
\floatcaption{VAw0}
{{\it Standard $\c^2$ fits to Eq.~(\ref{cauchy}) for $w(s)=1$, combined V\&A
channels. FOPT results are shown above the 
double horizontal line, CIPT results below.
The first line for each $s_{min}$ gives the V DV parameters; the second line
the A ones.  Every third value of $s_0$ 
starting at $s_{min}$ is included in the fits.
Errors are standard $\c^2$ errors; $\g_{V,A}$ and $\b_{V,A}$ in 
{\rm GeV}$^{-2}$.}}
\end{quotation}
\vspace*{-4ex}
\end{table}%
\clearpage

\begin{figure}[t]
\centering
\includegraphics[width=2.9in]{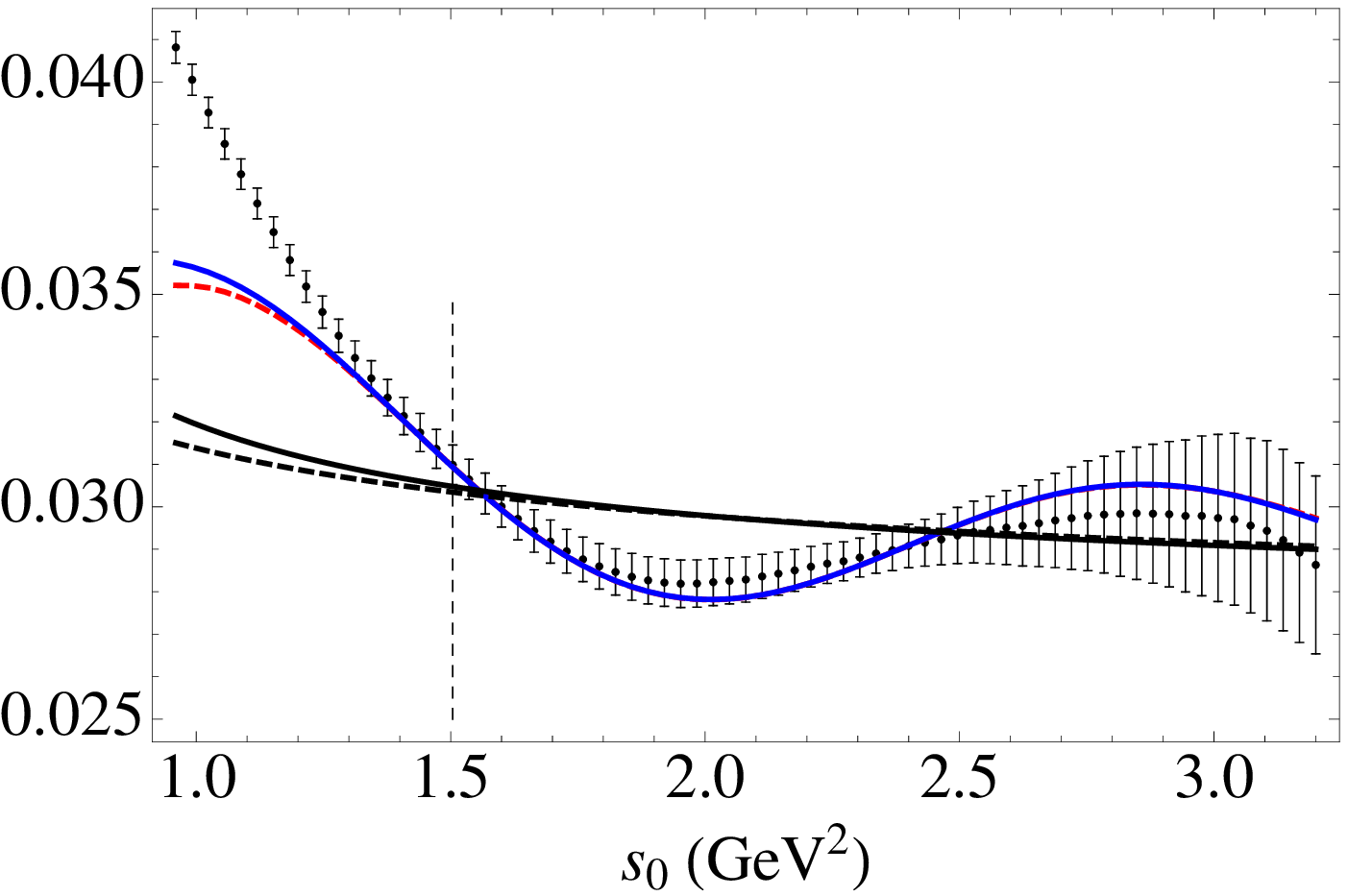}
\hspace{.1cm}
\includegraphics[width=2.9in]{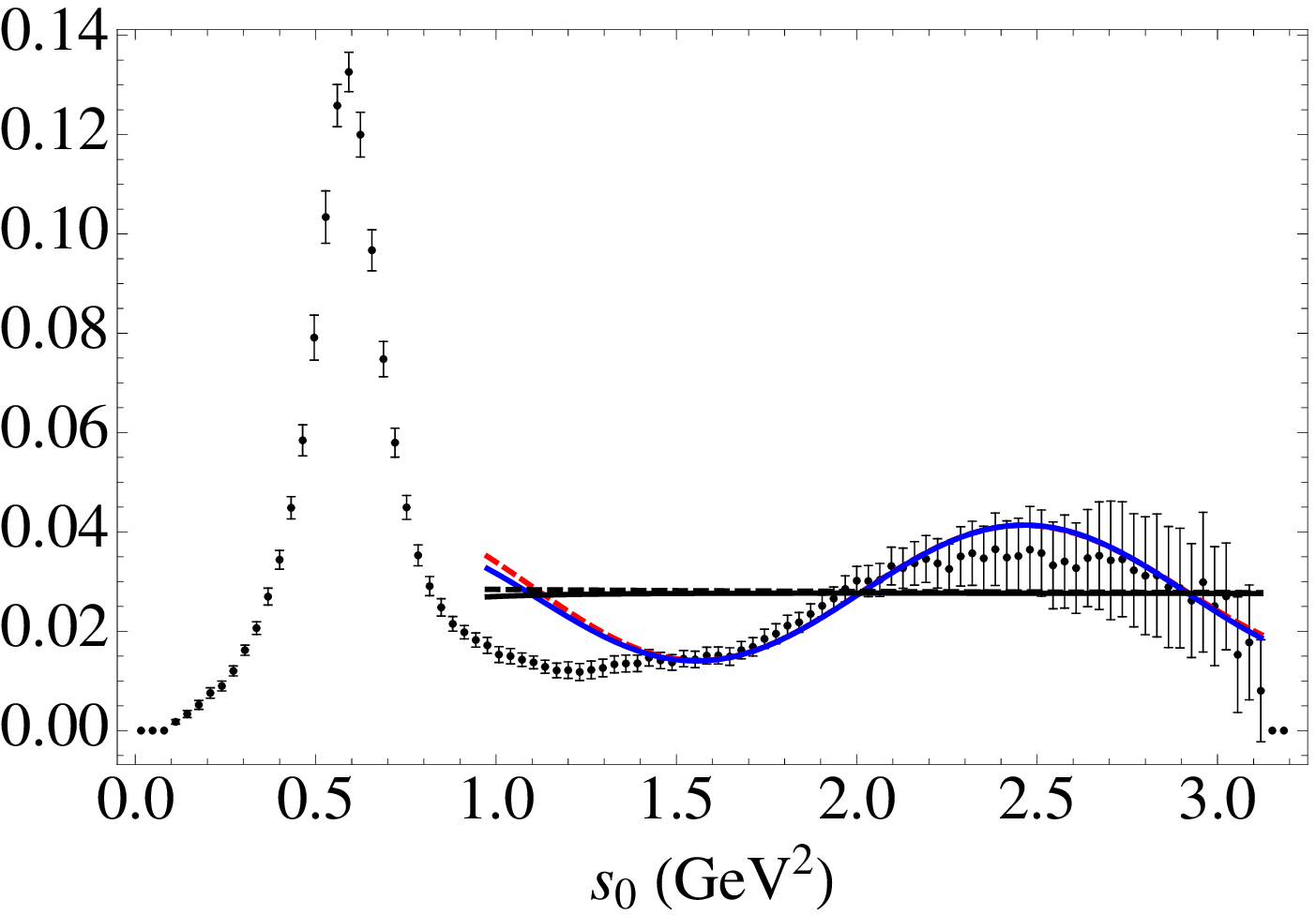}
\includegraphics[width=2.9in]{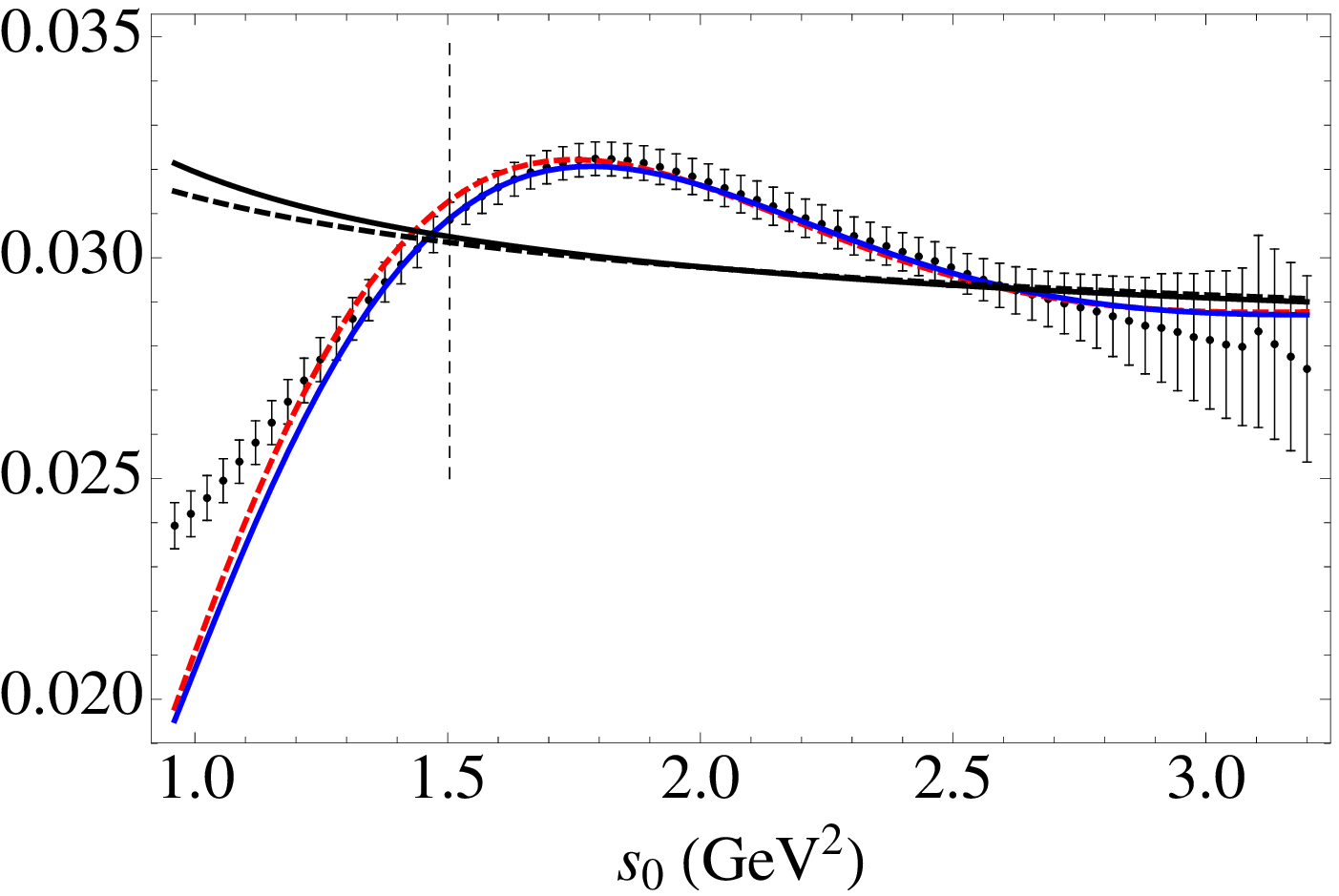}
\hspace{.1cm}
\includegraphics[width=2.9in]{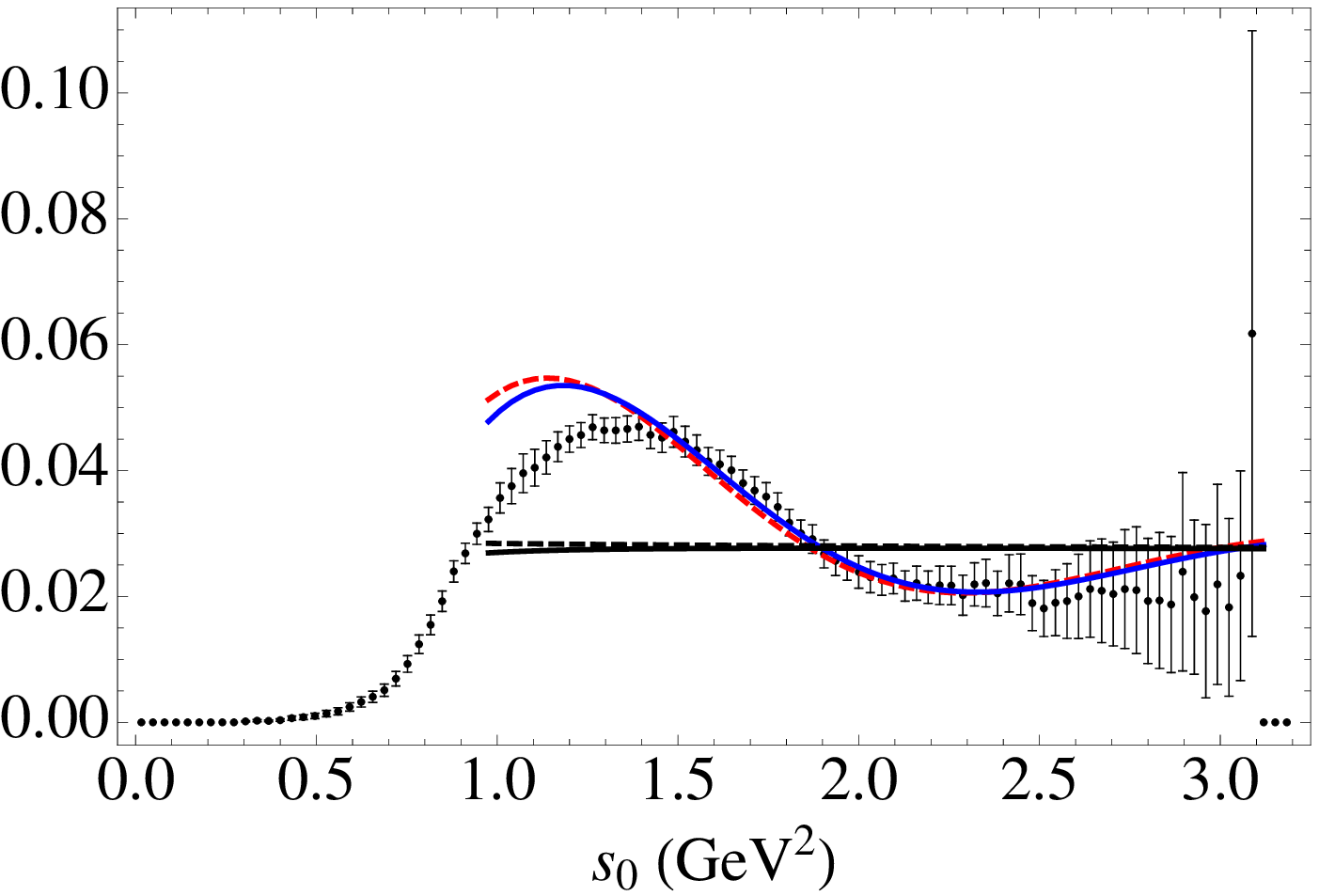}
\floatcaption{VAw0figure}{\it
Left panels: comparison of $I^{(\hw_0)}_{ex}(s_0)$ and $I^{(\hw_0)}_{th}(s_0)$
for the $s_{min}=1.5\ {\rm GeV}^2$ combined V$\&$ A fits of Table~\ref{VAw0}
(top: V channel, bottom: A channel). Right panels: comparison of the theoretical
spectral function resulting from this fit with the experimental results. CIPT
fits are shown in red (dashed) and FOPT in blue (solid). The (much flatter)
black curves display only the OPE parts of the FOPT (solid) and CIPT (dashed)
fit results. The vertical dashed lines indicate the location of $s_{min}$.}
\end{figure}

\vspace{0.9cm}
\begin{boldmath}
\subsection{\label{multiple} Multiple-weight fits with the weights $\hw_0$, $\hw_2$ and $\hw_3$}
\end{boldmath}
One would naively expect that by using more moments, more information could
be extracted from the data. This would help reducing the errors reported in
Tables~\ref{w0} and \ref{VAw0}. Higher-degree weights, however, also require
the introduction of additional OPE fit parameters. This, in combination with
the very strong correlations, may turn out to reduce the extra constraints
placed on the parameters ($\a_s$ and the DV parameters) entering the $\hat{w}_0$
FESR by the additional moments.  In addition, as we already pointed out in
Sec.~\ref{corr}, it appears to be impossible to perform standard $\chi^2$ fits
to multiple moments.

Table~\ref{w023} shows the results of simultaneous fits to moments with
weights $\hw_0$, $\hw_2$, and $\hw_3$ for the $V$ channel using the fit
quality~(\ref{blockcorr}), with errors computed from Eq.~(\ref{parcorr}); 
see Fig.~\ref{Vw023figure} for a visual representation of the qualities
of the resulting fits for the case $s_{min}=1.5$~GeV$^2$. Results are
consistent with those presented in Sec.~\ref{w=1}, but errors are slightly
larger.   

\begin{table}[thb]
\begin{center}
\vspace{0.5cm}
\begin{tabular}{|c|c|c|a|k|g|g|g|g|g|}
\hline
$s_{min}$ & dof & ${\cal Q}^2$/dof & \multicolumn{1}{c|}{$\alpha_s$} & \multicolumn{1}{c|}{$\kappa_V$} & \multicolumn{1}{c|}{$\gamma_V$} & \multicolumn{1}{c|}{$\alpha_V$} & \multicolumn{1}{c|}{$\beta_V$} & \multicolumn{1}{c|}{$10^2C_{6,V}$} & \multicolumn{1}{c|}{$10^2C_{8,V}$} \\
\hline
1.3 & 167 & 0.42 & 0.300(18) & 0.050(35) & 0.87(48) & 0.38(77) & 2.87(44) & -0.39(40) & 0.45(68)\\
1.4 & 158 & 0.33 & 0.304(17) & 0.027(18) & 0.46(43) & -0.48(88) & 3.35(48) & -0.43(31) & 0.67(47)\\
1.5 & 149 & 0.33 & 0.304(19) & 0.021(12) & 0.31(38) & -0.7(1.1) & 3.46(58) & -0.46(33) & 0.76(51)\\
1.6 & 140 & 0.33 & 0.305(23) & 0.025(17) & 0.41(43) & -0.6(1.4) & 3.41(74) & -0.43(46) & 0.68(76)\\
1.7 & 131 & 0.34 & 0.303(25) & 0.0136(95) & 0.10(39) & -0.8(1.5) & 3.47(73) & -0.50(45) & 0.88(71)\\
\hline
\hline
1.3 & 167 & 0.40 & 0.332(47) & 0.035(32) & 0.60(64) & 0.5(1.0) & 2.84(52) & -0.27(59) & 0.19(95)\\
1.4 & 158 & 0.32 & 0.327(31) & 0.023(16) & 0.34(46) & -0.3(1.0) & 3.26(54) & -0.43(36) & 0.58(58)\\
1.5 & 149 & 0.32 & 0.322(31) & 0.020(13) & 0.26(42) & -0.6(1.3) & 3.39(66) & -0.50(37) & 0.73(62)\\
1.6 & 140 & 0.33 & 0.323(42) & 0.025(17) & 0.37(47) & -0.5(1.7) & 3.35(89) & -0.48(54) & 0.66(98)\\
1.7 & 131 & 0.34 & 0.319(39) & 0.014(10) & 0.11(41) & -0.7(1.7) & 3.43(84) & -0.57(48) & 0.89(85)\\
\hline
\end{tabular}
\end{center}
\begin{quotation}
\floatcaption{w023}{{\it Fits to Eq.~(\ref{cauchy}) with weights 
$\hw_{0,2,3}$, V channel, using fit quality~(\ref{blockcorr}). 
FOPT results are shown above the double horizontal line,
CIPT fits below. Errors have been computed using Eq.~(\ref{parcorr});
$\g_V$ and $\b_V$ in {\rm GeV}$^{-2}$, $C_{6,V}$ in {\rm GeV}$^6$ and
$C_{8,V}$ in {\rm GeV}$^8$.}}
\end{quotation}
\vspace*{-4ex}
\end{table}%

\begin{figure}[thb]
\centering
\includegraphics[width=2.9in]{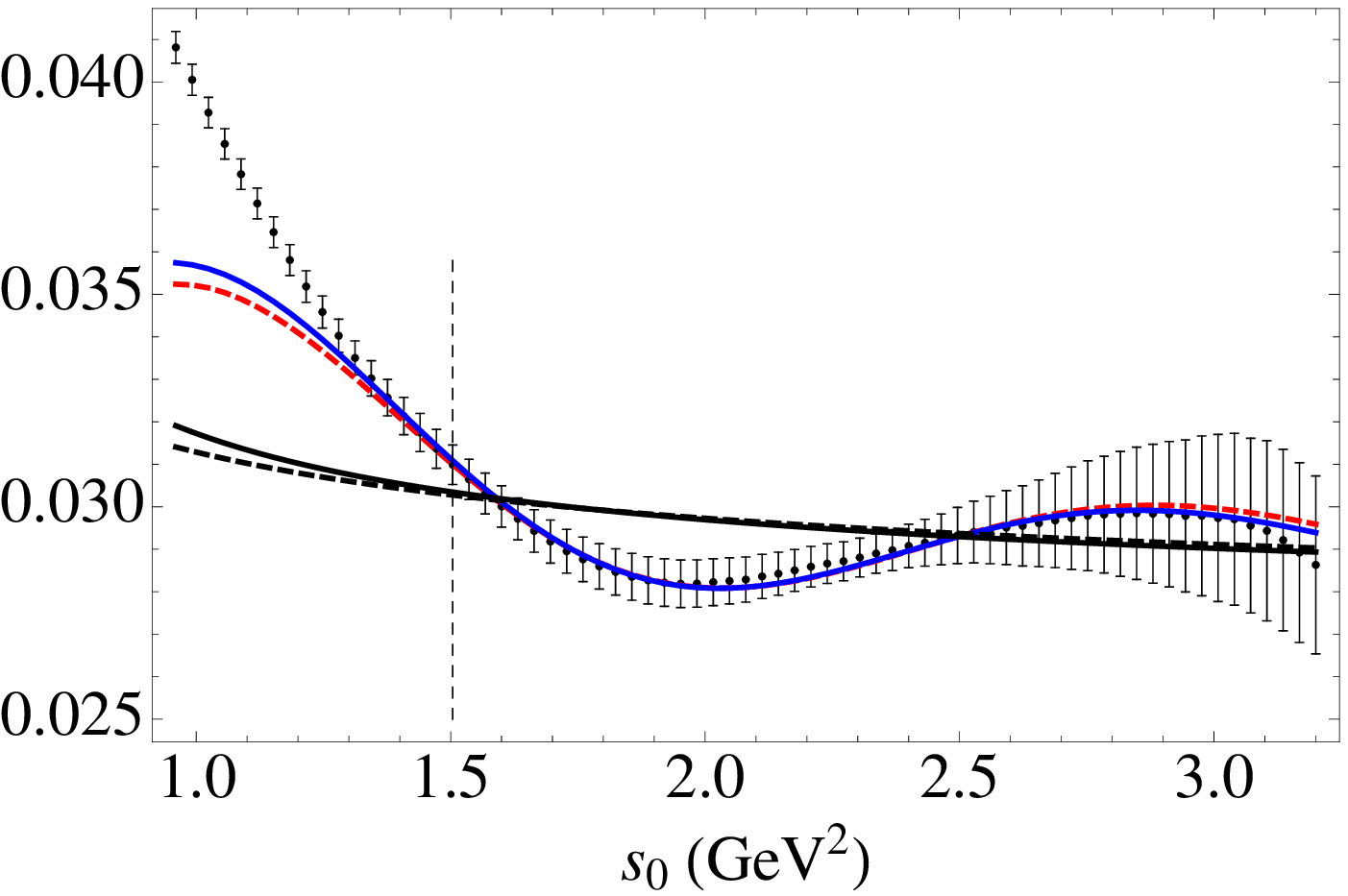}
\hspace{.1cm}
\includegraphics[width=2.9in]{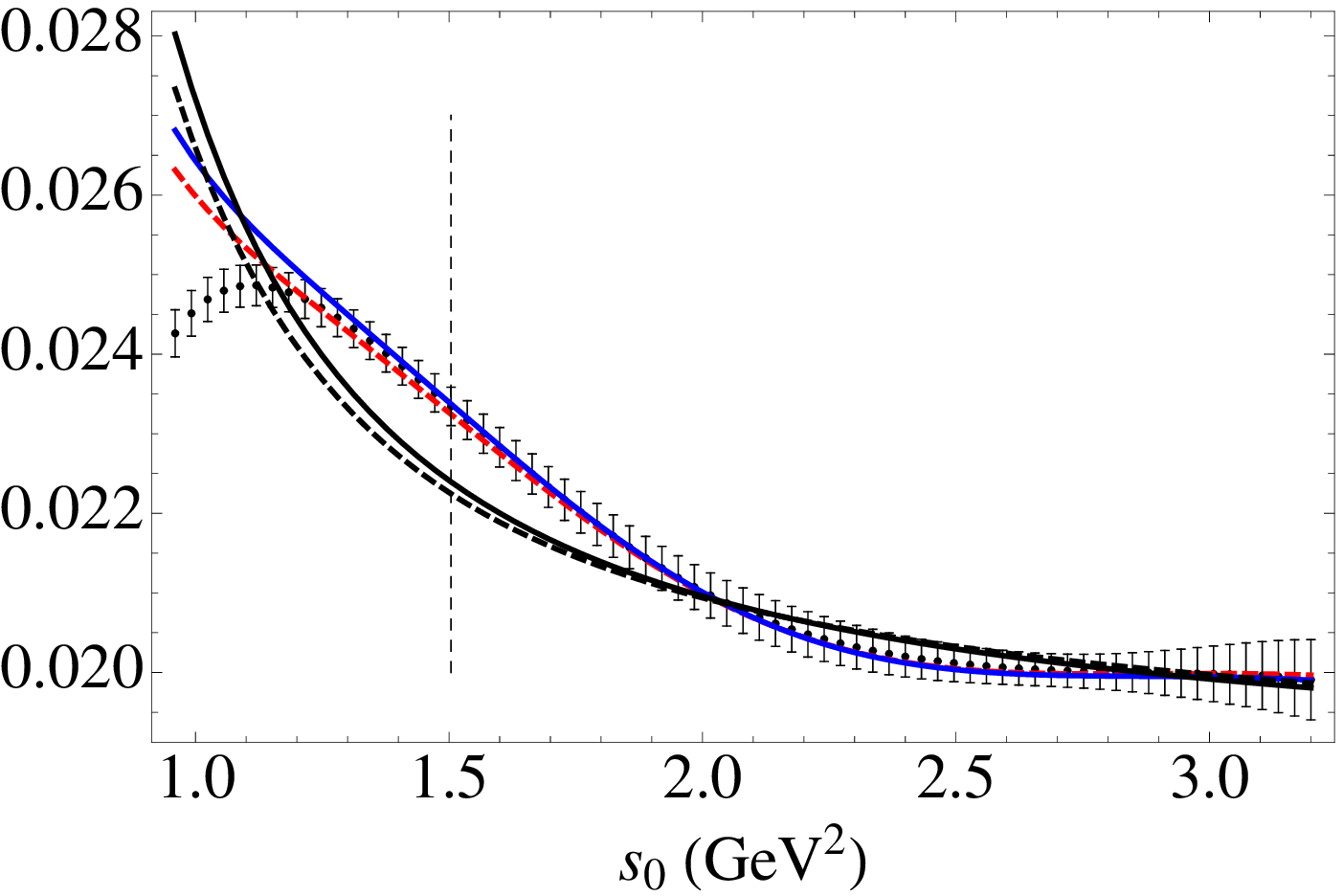}
\includegraphics[width=2.9in]{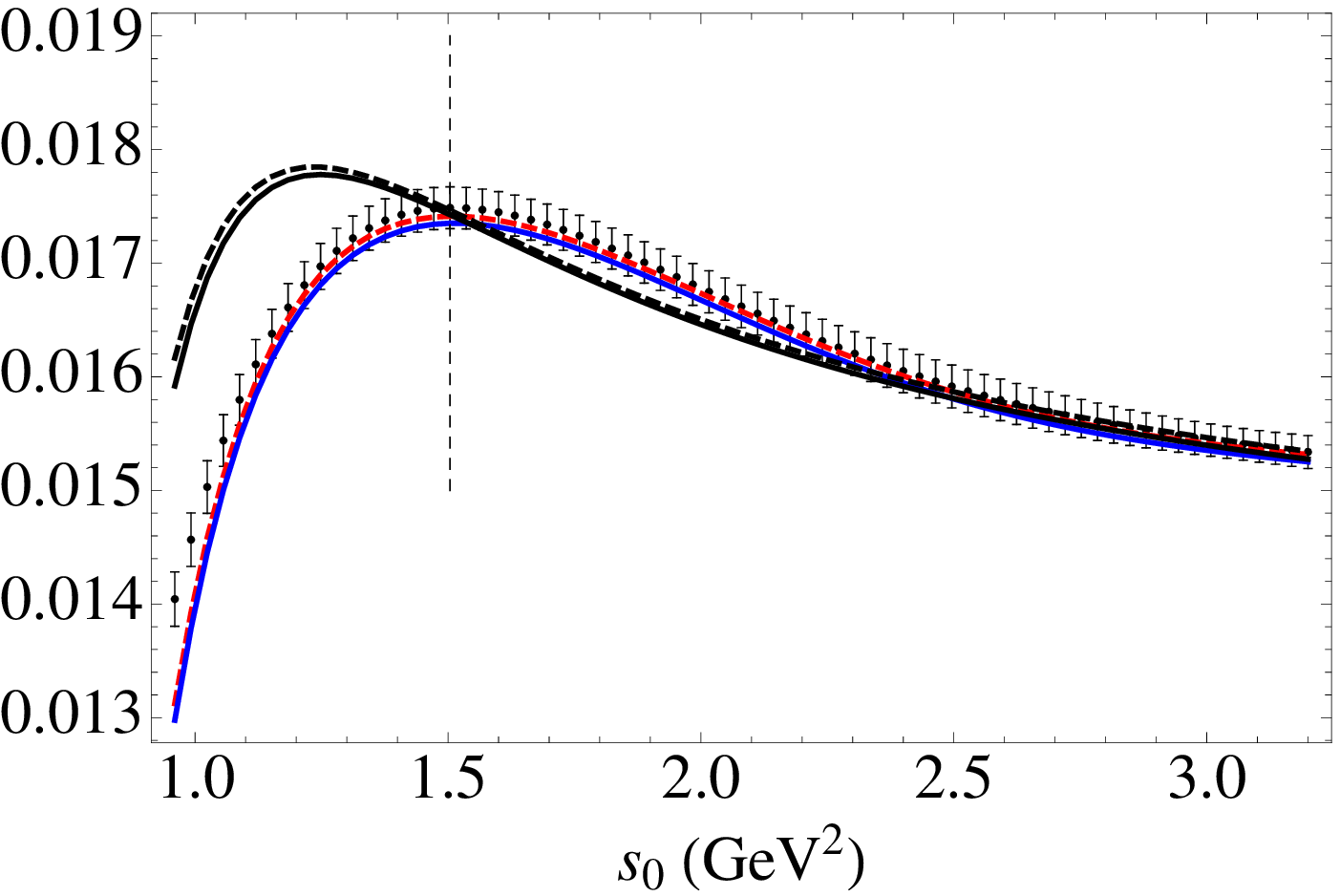}
\hspace{.1cm}
\includegraphics[width=2.9in]{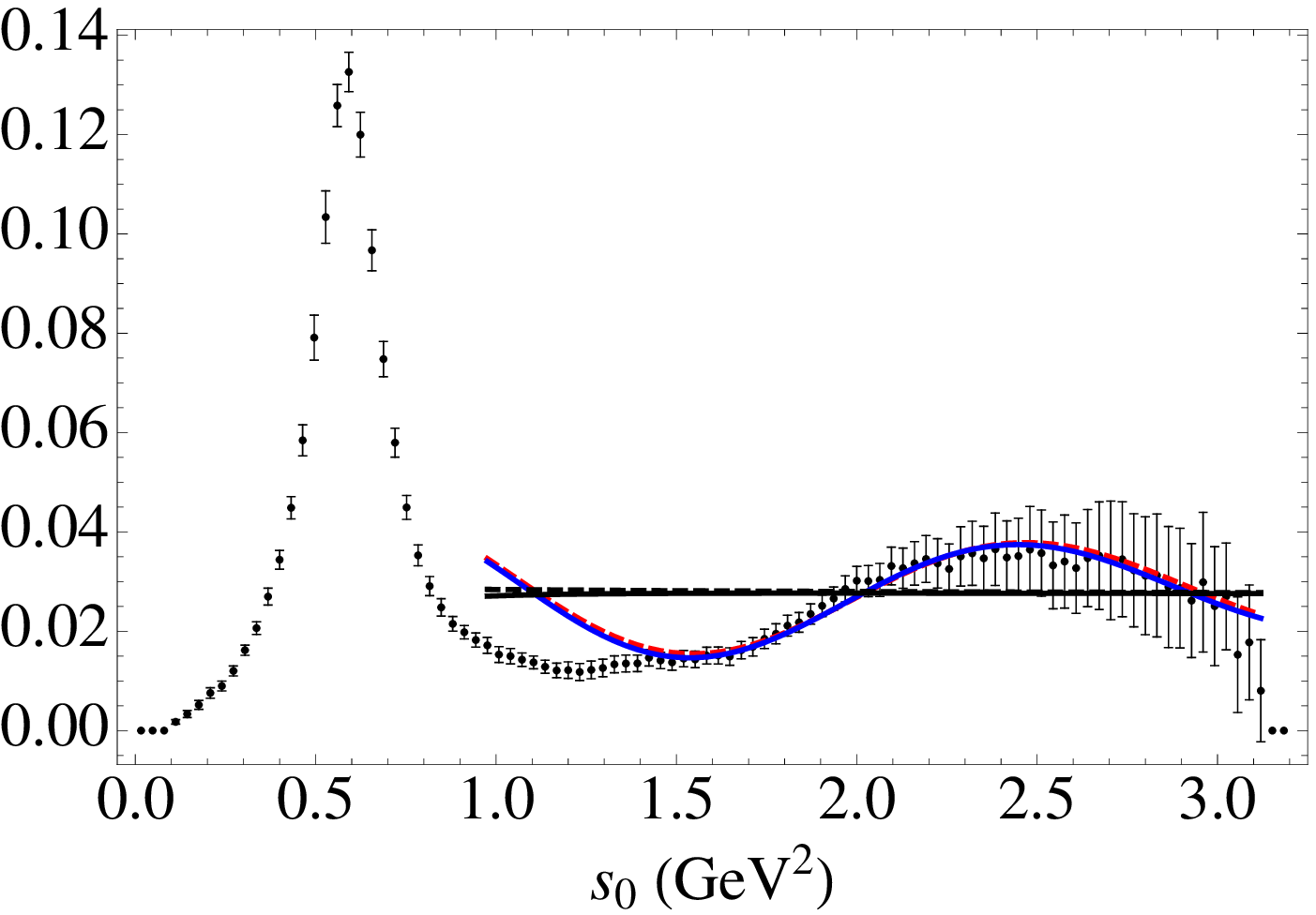}
\floatcaption{Vw023figure}{\it $V$-channel
fits of  Table~\ref{w023}, showing the theoretical and experimental versions of
the moments
$I^{(\hw_0)}$ (top left), $I^{(\hw_2)}$ (top right) and $I^{(\hw_3)}$ (bottom
left), for $s_{min}=1.5\ {\rm GeV}^2$. Bottom-right panel: comparison of the
theoretical spectral function resulting from this fit with the experimental
results. CIPT fits are shown in red (dashed) and FOPT in blue (solid). The
(flatter) black curves display only the OPE parts of the FOPT (solid) and CIPT
(dashed) fit results. The vertical dashed lines indicate the location of
$s_{min}$.}
\end{figure}

Again, the black curves in Fig.~\ref{Vw023figure} show the OPE parts of the 
theoretical curves, \ie, the results obtained by removing the DV contributions 
from the blue and red curve results. For the spectral function, 
$I_{ex}^{(\hw_0)}$ and $I_{ex}^{(\hw_2)}$, it is again clear that no good 
description of the data can be obtained without a model for DVs. This is not 
the case for the doubly-pinched moment $I_{ex}^{(\hw_3)}$. In this case, one 
would expect that a reasonably good fit can be obtained without DVs, for values of $s_{min}$ down to somewhere below $\sim 2$~GeV$^2$. This is consistent with
the results of Ref.~\cite{MY} for various doubly pinched weights, and, for the 
doubly pinched kinematic weight, also with the results of 
Refs.~\cite{ALEPH,OPAL}; in those cases, reasonably good matches for the sum 
of the vector and axial channels were obtained without the inclusion of DVs.
However, one would also expect that a best fit can only be obtained by 
{\em shifting} the OPE parameters relative to those reported in 
Table~\ref{w023}.

An estimate similar to Eq.~(\ref{asw0}), using the fit with 
$s_{min}=1.5$~GeV$^2$, leads to 
\begin{eqnarray}
\label{asw023}
\a_s(m_\t^2)&=&0.304\pm 0.019\pm 0.001\qquad\mbox{(FOPT)}\ ,\\
\a_s(m_\t^2)&=&0.322\pm 0.031\pm 0.005\qquad\mbox{(CIPT)}\ ,\nonumber
\end{eqnarray}
which is in excellent agreement with Eq.~(\ref{asw0}). We see, however, that the
simultaneous fit to multiple moments does not help reduce the error. 
It is an interesting question whether
fit qualities other than those of Eqs.~(\ref{diag}) and~(\ref{blockcorr}) exist that would
lead to smaller errors. Fits like those reported in Table~\ref{w023} 
using fit quality~(\ref{diag}) lead to consistent results, but with 
significantly larger errors than those shown in the table.

Finally, in Table~\ref{VAw023}, we show the results of
combined $V\& A$ channel fits, similar to those of 
Table~\ref{VAw0}, but for the
weights $\hw_0$, $\hw_2$ and $\hw_3$. {}From this table, we obtain
\begin{eqnarray}
\label{asw023VA}
\a_s(m_\t^2)&=&0.302\pm 0.015\pm 0.001\qquad\mbox{(FOPT)}\ ,\\
\a_s(m_\t^2)&=&0.322\pm 0.024\pm 0.008\qquad\mbox{(CIPT)}\ ,\nonumber
\end{eqnarray}
with the second error again reflecting the variation over the
range $s_{min}=1.4$ to $1.6$ GeV$^2$.  Comparing Table~\ref{VAw023} with 
Table~\ref{w023} we see that adding the $A$ channel gives
little extra information (apart from estimates of the axial OPE and DV 
parameters).   Adding the $A$ channel increases the central values of $\g_V$,
but all parameter values are consistent between these two tables within (sometimes
substantial) errors.  The fit with $s_{min}=1.7$~GeV$^2$ is clearly not meaningful
(in particular for FOPT),
and the errors indicate that at this value of $s_{min}$, $\cq^2$ is very flat in some directions
in parameter space.  Indeed, restricting the value of $\a_s$ to the value obtained
at $s_{min}= 1.6$~GeV$^2$ leads to a good fit for the remaining parameters,
consistent with the unrestricted $s_{min}= 1.6$~GeV$^2$ fit, and with a value for
$\cq^2$ almost equal to the value reported in the table.  

\begin{table}[t]
\begin{center}
\begin{tabular}{|c|c|c|a|k|g|g|g|g|g|}
\hline
$s_{min}$ & dof & ${\cal Q}^2$/dof & \multicolumn{1}{c|}{$\alpha_s$} & \multicolumn{1}{c|}{$\kappa_{V,A}$} & \multicolumn{1}{c|}{$\gamma_{V,A}$} & \multicolumn{1}{c|}{$\alpha_{V,A}$} & \multicolumn{1}{c|}{$\beta_{V,A}$} & \multicolumn{1}{c|}{$10^2C_{6,V/A}$} & \multicolumn{1}{c|}{$10^2C_{8,V/A}$} \\
\hline
1.3 & 104 & 0.66 & 0.305(11) & 0.033(18) & 0.57(37) & -0.11(63) & 3.15(37) & -0.36(24) & 0.51(38) \\
&&&& 0.053(20) & 0.78(23) & -0.93(63) & -2.81(36) & 0.08(27) & 0.26(52) \\
\hline
1.4& 98 & 0.48 & 0.303(13) & 0.023(12) & 0.32(37) & -0.76(77) & 3.51(43) & -0.45(23) & 0.74(35) \\
&&&& 0.082(41) & 0.98(29) & -1.30(75) & -2.61(42) & -0.12(41) & 0.78(91) \\
\hline
1.5 & 92 & 0.47 & 0.302(15) & 0.020(10) & 0.24(35) & -0.97(94) & 3.61(51) & -0.49(25) & 0.82(40) \\
&&&& 0.109(84) & 1.11(40) & -1.50(89) & -2.51(49) & -0.24(53) & 1.1(1.3) \\
\hline
1.6 & 86 & 0.46 & 0.302(19) & 0.024(16) & 0.35(43) & -0.9(1.2) & 3.58(66) & -0.48(36) & 0.78(60)\\
&&&& 0.19(20) & 1.37(52) & -1.7(1.2) & -2.44(63) & -0.36(74) & 1.6(2.0) \\
\hline
1.7 & 80 & 0.41 & 0.277(35) & 0.6(1.9) & 2.3(2.1) & -1.6(5.9) & 3.9(3.1) & -1.00(85) & 1.7(1.9)\\
&&&& 1.0(2.3) & 2.02(84) & -3.1(2.6) & -1.7(1.4) & -1.6(1.9) & 5.7(7.0) \\
\hline
\hline
1.3 & 104 & 0.55 & 0.348(19) & 0.0206(86) & 0.22(30) & 0.26(57) & 2.98(32) & -0.20(23) & 0.19(34) \\
&&&& 0.073(33) & 1.02(30) & -0.14(73) & -3.41(41) & 0.41(23) & -0.49(43) \\
\hline
1.4 & 98 & 0.45 & 0.330(22) & 0.0188(96) & 0.18(35) & -0.48(82) & 3.36(45) & -0.43(25) & 0.61(40) \\
&&&& 0.085(42) & 1.03(29) & -0.84(93) & -2.87(52) & 0.04(40) & 0.30(86) \\
\hline
1.5 &92 & 0.45 & 0.322(24) & 0.0185(95) & 0.18(34) & -0.8(1.0) & 3.51(55) & -0.52(28) & 0.77(47) \\
&&&& 0.105(80) & 1.11(40) & -1.2(1.0) & -2.66(58) & -0.15(55) & 0.8(1.3) \\
\hline
1.6 & 86 & 0.41 & 0.321(31) & 0.023(15) & 0.31(41) & -0.7(1.4) & 3.49(73) & -0.51(40) & 0.73(73)\\
&&&& 0.18(19) & 1.36(52) & -1.4(1.4) & -2.58(76) & -0.28(80) & 1.2(2.1) \\
\hline
1.7 & 80 & 0.42 & 0.317(37) & 0.022(17) & 0.29(43) & -0.8(1.7) & 3.54(86) & -0.55(51) & 0.81(96)\\
& & & & 1.0(2.2) & 2.13(88) & -2.2(2.7) & -2.1(1.4) & -0.7(1.3) & 2.7(4.4) \\
\hline
\end{tabular}
\end{center}
\begin{quotation}
\floatcaption{VAw023}{{\it Fits to Eq.~(\ref{cauchy}) with 
weights $\hw_{0,2,3}$, combined $V$ and
$A$ channels, using fit quality~(\ref{blockcorr}). FOPT results are shown above 
the double horizontal line, CIPT fits below.
Errors have been computed with Eq.~(\ref{parcorr});
$\g_{V,A}$ and $\b_{V,A}$ in {\rm GeV}$^{-2}$, $C_{6,V}$ and $C_{6,A}$ in 
{\rm GeV}$^6$ and
$C_{8,V}$ and $C_{8,A}$ in {\rm GeV}$^8$.
The first line for each $s_{min}$ gives the $V$ channel OPE and DV parameters; 
the second line the $A$ channel ones.
Every third value of $s_0$ starting at $s_{min}$ 
is included in the fits.}}
\end{quotation}
\vspace*{-4ex}
\end{table}%

Let us next deduce some implications of the above fits for the breaking of the
factorization hypothesis in the $D=6$ condensates. To begin with, we will assume
that the $D=6$ condensates are dominated by their leading-order contribution. The
corresponding contribution to the $V/A$ correlators is given by \cite{BNP}:
\begin{eqnarray}
\label{PiD6OP}
C_{6,V/A} &\!\!=\!\!& -8\,\pi^2 a_s\,
\Bigl\langle (\bar u\gamma_\mu \!\left(\!\!\begin{array}{c} \gamma_5 \\
{1} \end{array}\!\!\right)\! t^a d) (\bar d\gamma^\mu\!\left(\!\!
\begin{array}{c} \gamma_5 \\ {1} \end{array}\!\!\right)\!t^a u) \Bigr\rangle
 \\
\vbox{\vskip 8mm}
&& \hspace{-2mm} -\,\frac{8}{9}\,\pi^2 a_s\, \sum\limits_{q=u,d,s} \left\langle(\bar u
\gamma_\mu t^a u + \bar d\gamma_\mu t^a d)(\bar q\gamma^\mu t^a q)
\right\rangle + \co(a_s^2) \ ,\nonumber
\end{eqnarray}
where, in the first line, the upper Dirac structure $\g_5$ corresponds to the $V$ channel
and the lower $1$ to the $A$ channel. 
The two four-quark
condensates can be parametrized in terms of their factorization values
by
\begin{eqnarray}
\label{rho1rho5}
\langle(\bar q_i\gamma_\mu t^a q_j)(\bar q_j\gamma^\mu t^a q_i)\rangle
&\!\!\equiv\!\!& -\,\frac{4}{9}\,\langle\bar q_i q_i\rangle
\langle\bar q_j q_j\rangle\!\cdot\!\rho_1 \ ,  \\
\vbox{\vskip 6mm}
\langle(\bar q_i\gamma_\mu\gamma_5 t^a q_j)(\bar q_j\gamma^\mu\gamma_5 t^a q_i)
\rangle &\!\!\equiv\!\!& \phantom{-}\,\frac{4}{9}\,\langle\bar q_i q_i\rangle
\langle\bar q_j q_j\rangle\!\cdot\!\rho_5 \ ,\nonumber
\end{eqnarray}
where the parameters $\r_1$ and $\r_5$ would be equal to one if
the vacuum-saturation approximation were exact. Further assuming that isospin
breaking is small in the light $u$- and $d$-quark sector, that is
$\langle\bar uu\rangle=\langle\bar dd\rangle\equiv\langle\bar qq\rangle$,
Eqs.~(\ref{PiD6OP}) and~\ref{rho1rho5} imply
\begin{equation}
\label{C6VA}
C_{6,V/A} \,=\, \frac{32}{81}\,\pi^2 a_s\,
\langle\bar qq\rangle^2 \!\left(\!\!\begin{array}{c} 2\,\rho_1 -
9\,\rho_5 \\ 11\,\rho_1 \end{array}\!\!\right)\! \ .
\end{equation}
Inverting Eq.~(\ref{C6VA}), on the basis of the results of Table~\ref{VAw023}, estimates
of the parameters $\r_{1,5}$ can be deduced. As representative examples,
for the central fits with $s_{min}=1.5$~GeV$^2$, one obtains:
\begin{eqnarray}
\label{rho1rho5num}
\rho_1 &\!\!=\!\!& -\,1.4 \pm 3.2 \,, \quad
\rho_5 \,=\, 3.3 \pm 1.5 \qquad \mbox{(FOPT)} \ , \\
\rho_1 &\!\!=\!\!& -\,0.9 \pm 3.1 \,, \quad
\rho_5 \,=\, 3.4 \pm 1.5 \qquad \mbox{(CIPT)} \ ,\nonumber
\end{eqnarray}
where $\langle\bar qq\rangle(m_\tau^2)=-\,(272\,{\rm MeV})^3$ \cite{jam02},
together with our results for $\a_s(m_\t^2)$, has been employed. The
central results of Eq.~(\ref{rho1rho5num}) display sizable deviations from the
factorization values $\r_1=\r_5=1$, though, given the large uncertainties,
the significance is not very high. The employed perturbative resummation scheme
does not seem to play a big role for the condensate or DV parameters, suggesting
that this choice is mostly compensated for by the differing $\alpha_s$ values.\\

\subsection{\label{errors} Errors from truncating perturbation theory}
One of the uncertainties afflicting any determination of $\a_s$ is that 
perturbation theory needs to be truncated, irrespective of whether the
CIPT or FOPT resummation scheme is used for the truncated series.
As already mentioned in Sec.~\ref{theory}, we use the known values
of $c_{n1}$ up to $n=4$, together with the estimate $c_{51}=283$. 
We assign a $100\%$ error, $\pm 283$, to this estimate, using this
as a measure of the truncation uncertainty. 

For the fits presented in Eqs.~(\ref{asw0}),~(\ref{asw0VA}),~(\ref{asw023}) and~(\ref{asw023VA}) we find that this variation of $c_{51}$
leads to a shift of at most $\pm 0.006$ in $\a_s(m_\t^2)$; for Eq.~(\ref{asw0}) the 
shift is $\pm 0.005$. The
shifts in other parameters are also small (well within fitting errors). 

\begin{boldmath}
\subsection{\label{V+A} Consistency with the $V+A$ and $V-A$ chiral sum rules}
\end{boldmath}
While Figs.~\ref{Vw0figure}--\ref{Vw023figure} show that the parameter values
obtained from the fits corresponding to these figures give a good
description of the data, with those parameter values in hand one may also 
consider other quantities. The total non-strange scaled $V+A$ branching 
fraction $R_{V+A;ud}$ (\seef\ Eqs.~(\ref{defR}),~(\ref{taukinspectral}) and~(\ref{RVpA}))
has always played
a central role in the study of hadronic $\t$ decays.
In particular, following Refs.~\cite{ALEPH,OPAL},
we may consider $R_{V+A;ud}(s_0)$ for a hypothetical $\t$ of mass-squared
$m_\t^2=s_0$, as a function of $s_0$. We show our version of this quantity in
Fig.~\ref{Rtaufigure}, using the parameter values for the fit with 
$s_{min}=1.5$~GeV$^2$ of Table~\ref{VAw023}, the
only fit reported that simultaneously yields all parameters
needed to evaluate $R_{V+A;ud}(s_0)$.\footnote{We included the 
pion-pole contribution to the longitudinal part in Eq.~(\ref{taukinspectral}) in both the
data points and the theory curves in Fig.~\ref{Rtaufigure}.} Clearly, our result compares 
well with the fits shown in Fig.~10 of Ref.~\cite{OPAL},\footnote{For an early investigation of 
this type, see Ref.~\cite{GN}.} especially keeping in mind 
that in Fig.~\ref{Rtaufigure} we only show errors on the experimental 
spectral integrals, and not on the theory curves. 
Both CIPT and FOPT describe the data well
down to $s_0=1.5$~GeV$^2$. We used the same values for $S_{EW}=1.0194$
and $|V_{ud}|^2=0.9512$ as Ref.~\cite{OPAL} in order to plot $R_{V+A;ud}(s_0)$.
In view of the discussion in Ref.~\cite{MY} of the analysis of Refs.~\cite{ALEPH,OPAL},
we conclude that our fits pass the test of $R_{V+A;ud}(s_0)$ much better than
the original analyses of Refs.~\cite{ALEPH,OPAL}, and over a wider range of $s_0$ than the alternate, self-consistent
fits obtained ignoring DVs in Ref.~\cite{MY}. We emphasize, though, that
it is not sufficient to find a satisfactory description of this quantity 
only -- at the
very least all FESRs used in the fits should show a similarly good 
match between experiment and theory as a function of $s_0$,
as was shown to be the case for our
fits in Secs.~\ref{w=1} and \ref{multiple} above.

\begin{figure}[t]
\centering
\includegraphics[width=4in]{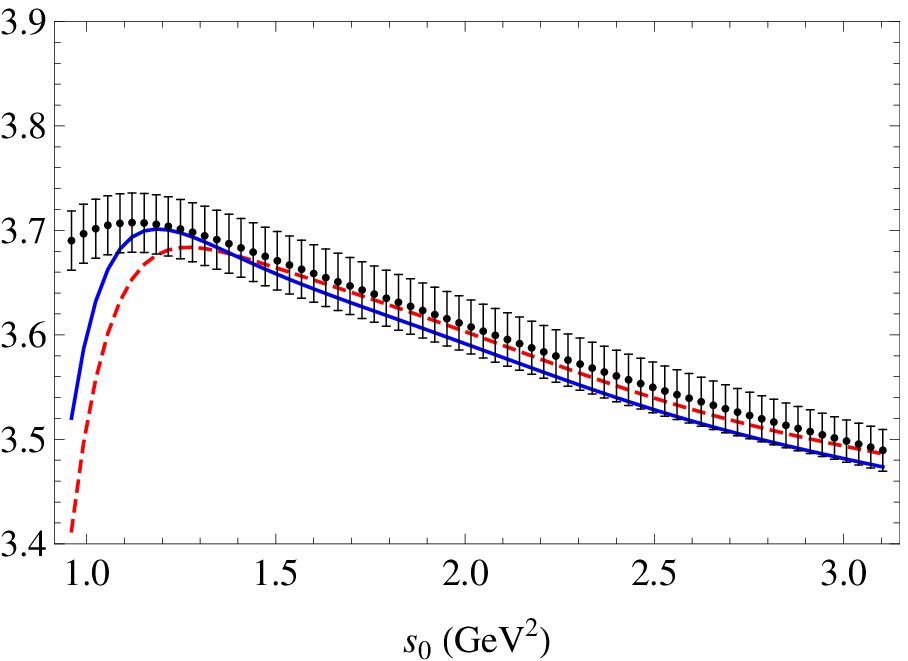}
\floatcaption{Rtaufigure}{\it $R_{V+A;ud}(s_0)$ as a function of $s_0$, with
$s_{min}=1.5$~{\rm GeV}$^2$ theory curves from 
Table~\ref{VAw023} for CIPT (dashed red curve) and
FOPT (solid blue curve).}
\end{figure}

We may perform a similar test on the fits reported in 
Table~\ref{w023}.
Of course, in that case only $V$-channel parameters are available, so one 
should consider the corresponding ratio $R_{V;ud}$ as a function of $s_0$. 
For the $V$ channel, $R_{V;ud}(s_0)$ coincides (up to the overall factor 
$12\p^2 S_{EW}|V_{ud}|^2$) with $I_{ex}^{(\hw_3)}(s_0)$. 
The corresponding match between data and theory
for $s_{min}=1.5$~GeV$^2$ is shown in the left bottom panel of
Fig.~\ref{Vw023figure}.

\begin{figure}[t]
\centering
\includegraphics[width=2.9in]{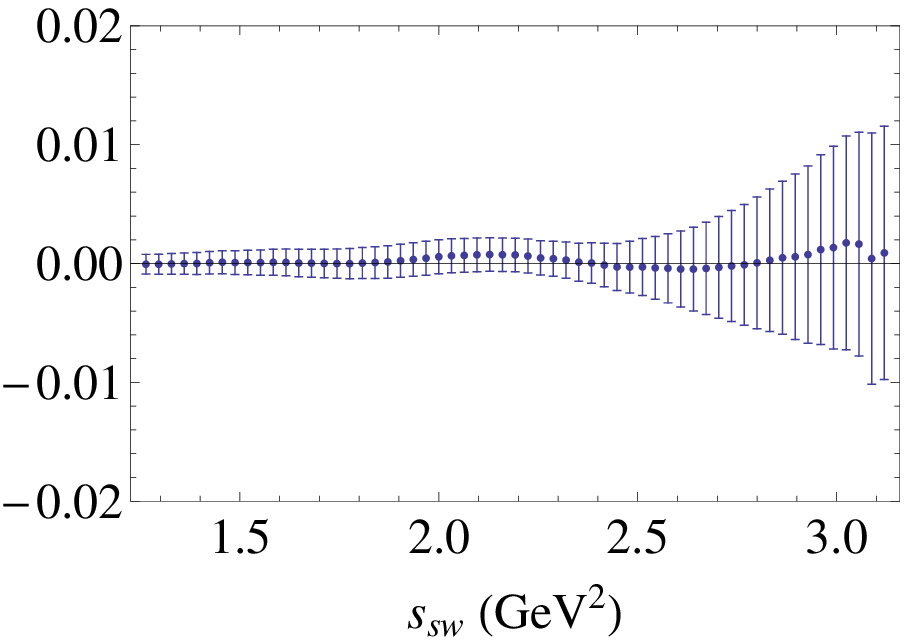}
\hspace{.1cm}
\includegraphics[width=2.9in]{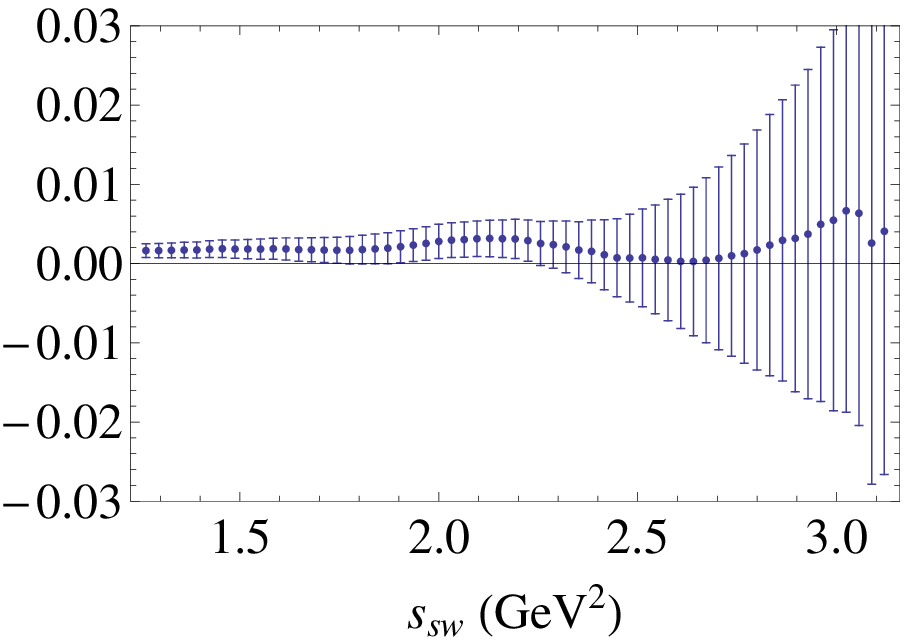}
\floatcaption{WSRfigure}{\it First (left, in {\rm GeV}$^2$) and second (right, 
in {\rm GeV}$^4$) Weinberg 
sum rules. Data are used for $s\le s_{sw}$, while the 
DV \ansatz~(\ref{ansatz}) with values from Table~\ref{VAw023} 
for $s_{min}=1.5~{\rm GeV}^2$
has been used for $s\ge s_{sw}$. FOPT figures are shown; CIPT figures look
essentially identical.}
\end{figure}

We can also test our results by considering how well they satisfy
the classical chiral $V-A$ sum rule constraints represented by the
two Weinberg sum rules \cite{SW} and the DGMLY sum rule
for the $\pi$ electromagnetic mass splitting \cite{EMpion}. These tests focus
specifically on the DV part of our spectral functions, since the 
OPE for the $V-A$ correlator (and hence the OPE part of the
spectral {\it ansatz}) has no $D=0$ contribution
and $D\ge 2$ OPE contributions to $\r_{V/A}(s)$ are tiny, and can be ignored. 

Weinberg's sum rules can be written as
\begin{eqnarray}
\label{WSR}
\int_0^\infty ds\left(\r^{(1+0)}_V(s)-\r^{(1+0)}_A(s)\right)&=&
\int_0^\infty ds\left(\r^{(1)}_V(s)-\r^{(1)}_A(s)\right)-2f_\p^2=0\ ,\\
\int_0^\infty ds\,s\left(\r^{(1+0)}_V(s)-\r^{(1+0)}_A(s)\right)&=&
\int_0^\infty ds\,s\left(\r^{(1)}_V(s)-\r^{(1)}_A(s)\right)-2m_\p^2 f_\p^2=0\ ,\nonumber
\end{eqnarray}
where we assumed, as before, that we can neglect terms of order $m_im_j$
with $i,j=u,d$, even though there is a term of order $m_im_j\a_s^2$ linearly
divergent in $s_0$ in the second sum rule.  The fact that this term is still 
very small at $s_0=m_\t^2$ amounts to the observation that the 
chiral symmetry breaking terms in the second of Eq.~(\ref{WSR}) are not visible 
in the data. This means that in our test of the second sum rule we can 
assume ourselves to be effectively in the chiral limit, in which the
divergence does not appear. In Fig.~\ref{WSRfigure} we show the 
integrals in Eq.~(\ref{WSR}) as a function of the ``switch point'' $s_{sw}$
below which experimental spectral data is used and above which
the difference of the $V$ and $A$
DV {\it ans\"atze} of Eq.~(\ref{ansatz}) is employed for $\rho_{V-A}(s)$.
The DV contributions were obtained using the DV parameter values of 
Table~\ref{VAw023}. If the DV \ansatz\ is, as we have
assumed, reliable in the window of $s_0$ employed in the FESR
fits which produce these DV parameter values, the $V-A$ sum rules
should be satisfied for all values of $s_{sw}$ lying in this
$s_0$ fit window. We see that this condition is well satisfied
for both of the Weinberg sum rules. 
The errors shown are those from the experimental ($s<s_{sw}$) part of the
integral on the left-hand side of the sum rules only. 

The first Weinberg sum rule is, of course, closely related to the difference
of the $V$ and $A$ $\hw_0(x)=1$ FESRs.  Therefore, the very good quality
of the matches between the experimental spectral integrals and
our theoretical representations thereof precludes
the first Weinberg sum rule being badly broken by our fits.
The second Weinberg
sum rule (as well as the DGMLY sum rule discussed below) can be viewed as
a prediction, because the fits of Table~\ref{VAw023} do not involve
any weight with a term linear in $s$.
We also note that if indeed we take the second sum rule in the 
chiral limit, we should omit the term $-2m_\p^2 f_\p^2$. 
The value of this term is $-0.00034$~GeV$^4$, and it can thus indeed
safely be dropped from the sum rule --- the difference would not be 
visible in the figure.

\begin{figure}[t]
\centering
\includegraphics[width=4.0in,height=3.0in]{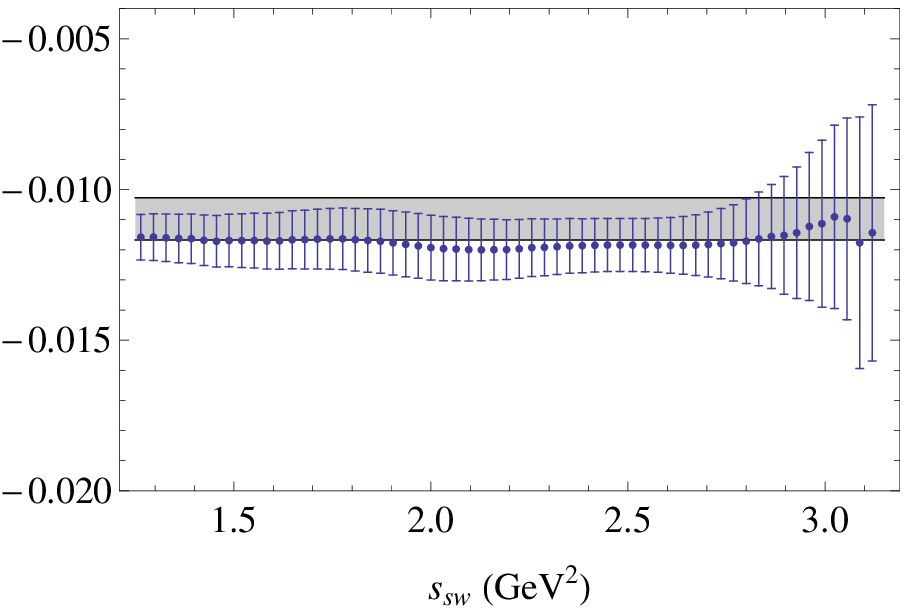}
\floatcaption{WSR3figure}{\it The DGMLY $\pi$ electromagnetic
mass difference sum rule:  left-hand side (data points) and right-hand side
(gray band) of 
Eq.~(\ref{pionmassdiff}), as a function of $s_{sw}$, in {\rm GeV}$^4$.
Data are used for $s\le s_{sw}$, the DV \ansatz~(\ref{ansatz}), with 
fit parameter values from the $s_{min}=1.5~{\rm GeV}^2$ entries
of Table~\ref{VAw023}, for $s\ge s_{sw}$. The FOPT 
figure is shown; the CIPT figure is essentially identical.
The gray band represents an estimate of the error on the right-hand
side of Eq.~(\ref{pionmassdiff}).}
\end{figure}

Finally, we consider the DGMLY sum rule for the $\pi$ electromagnetic mass
difference. To leading order in the chiral expansion, one has that
\begin{equation}
\label{pionmassdiff}
\int_0^\infty ds\,s\log{(s/\m^2)}\left(\r^{(1)}_V(s)-\r^{(1)}_A(s)\right)=-\frac{8\p f_0^2}{3\a}
\left(m_{\p^\pm}^2-m_{\p^0}^2\right)\ ,
\end{equation}
where $\a$ is the fine-structure constant, and $f_0$ the $\pi$ decay constant
in the two-flavor chiral limit.   On the right-hand side we take as input the values
$f_0=87.0\pm 0.6$~MeV \cite{MILC} and $m_{\p^+}^2-m_{\p^0}^2=0.00126\pm 0.00008$~GeV$^2$
\cite{PDG}.\footnote{Our uncertainty on the electromagnetic
contribution to the difference $m_{\p^+}^2-m_{\p^0}^2$ comes from
an estimate of the contribution of $m_u-m_d$ to the pion mass difference
\cite{ABT}.}
Because of the second Weinberg sum rule, the 
left-hand side of Eq.~(\ref{pionmassdiff}) is, in fact, independent of the scale 
$\m$. In Fig.~\ref{WSR3figure} we show the
left-hand  and the right-hand sides of Eq.~(\ref{pionmassdiff}), 
as a function of $s_{sw}$.   The left-hand side is represented by the data points,
while the gray band represents the right-hand side, including the error
resulting from the uncertainties in $f_0$ and the pion electromagnetic mass difference.
As for the Weinberg sum rules, 
the integral on the left-hand side is computed using
experimental data for $s\le s_{sw}$, and the DV \ansatz,
with DV parameters from the $s_{min}=1.5$~GeV$^2$ entries of 
Table~\ref{VAw023}, for $s\ge s_{sw}$. 
The errors shown again come from the experimental data part of the integral
only. We chose $\m^2=2.5$~GeV$^2$, which makes the errors in the figure 
relatively small. For lower values of $\m^2$ the errors are larger for 
larger $s_{sw}$, but the results are consistent with Eq.~(\ref{pionmassdiff}) being
satisfied for all $\m^2$ between $1.5$~GeV$^2$ and $m_\t^2$.

One might consider using these sum rules as a further
constraint on the DV parameters.\footnote{See,
for instance, Ref.~\cite{VAVal}, where, however, the {\it difference} $V-A$ 
was modeled with an expression of the form~(\ref{ansatz}).  
Given the results of
Tables~\ref{VAw0} and \ref{VAw023}, such a parametrization does not seem to be
favored by the data.
In the present work we avoid any additional
assumptions about the relations between the DV parameters in the
$V$ and $A$ channels by introducing separate DV parameters for each channel.}
Of course, this can only be done for combined $V\&A$ fits, 
which requires us to assume that the larger of the two $s_{min}$
values for the $V$ and $A$ channels still lies sufficiently
far below $m_\tau^2$ to make such a combined fit, using our DV
\ansatz\ in both channels, reliable. Since in this work our
primary results are obtained from purely $V$-channel fits, 
which do not require this assumption, we postpone 
this possibility to a future investigation.

\vspace{1.0cm}
\section{\label{summary} Summary of results}
Our most stable results come from fits to
only the $V$ channel. Moreover, since we need to include a model for DVs,
using only the $V$ channel avoids the additional assumption that our \ansatz\
already adequately describes the $A$-channel data in the interval between 
$1.5$~GeV$^2$ and $m_\t^2$. We emphasize, however, that all the results 
from our combined $V$ and $A$ channel fits are consistent with
those from our $V$-channel analyses.  In addition, they 
pass several further $V+A$ and $V-A$ channel tests, as 
demonstrated in Sec.~\ref{V+A}.

In view of these observations,
we will choose the fit to $I_{ex}^{(\hw_0)}$ in the vector channel as our
central result.    Adding the errors from the fit, the variation 
of $s_{min}$ and the variation of $c_{51}$ in quadrature, we find
\begin{eqnarray}
\label{finaltau}
\a_s(m_\t^2)&=&0.307\pm 0.019\qquad(\overline{MS},\ n_f=3,\ \mbox{FOPT})\ , \\
\a_s(m_\t^2)&=&0.322\pm 0.026\qquad(\overline{MS},\ n_f=3,\ \mbox{CIPT})\ .\nonumber
\end{eqnarray}
We do not average over CIPT and FOPT results, because 
we believe that it is useful to see the difference between values for 
$\a_s$ obtained with
the two resummation schemes after all non-perturbative
effects have been consistently taken into account. 
Running these values up to the $Z$ mass $M_Z$ yields \cite{CKS}\footnote{
The specifics of the evolution to the $Z$ mass are as
discussed in Ref.~\cite{BJ}.}
\begin{eqnarray}
\label{finalZ}
\a_s(M_Z^2)&=&0.1169\pm 0.0025\qquad(\overline{MS},\ n_f=5,\ \mbox{FOPT})\ , \\
\a_s(M_Z^2)&=&0.1187\pm 0.0032\qquad(\overline{MS},\ n_f=5,\ \mbox{CIPT})\ ,\nonumber
\end{eqnarray}
where we symmetrized the slightly asymmetric errors one obtains after running
up to the $Z$ mass.

These values can be compared to those obtained by OPAL from the same data
\cite{OPAL}.  The OPAL values at the $\t$ mass are
\begin{eqnarray}
\label{OPALtau}
\a_s(m_\t^2)&=&0.324\pm 0.014\qquad(\overline{MS},\ n_f=3,\ \mbox{FOPT,\ OPAL})\ , \\
\a_s(m_\t^2)&=&0.348\pm 0.021\qquad(\overline{MS},\ n_f=3,\ \mbox{CIPT,\ OPAL})\ ,\nonumber
\end{eqnarray}
where we added the experimental and theoretical errors quoted by OPAL
in quadrature.   We observe the shift to lower central values, with somewhat
larger errors that follow from using our new framework for analyzing the data.
We also note that our CIPT and FOPT values are somewhat closer, and that,
because of the larger errors, the difference between our two values 
for $\a_s$ is less significant.  

\section{\label{conclusion} Conclusion}
In this article, we provided a new framework for the extraction of 
$\a_s$ and other
OPE parameters from hadronic $\t$ decays. This new framework combines
two elements that have not been taken into account in the ``traditional''
analysis of hadronic $\t$ decays. One is a consistent treatment of the OPE, as
discussed in Ref.~\cite{MY}, and the other is a detailed quantitative estimate 
of the effect of violations of quark-hadron duality, using a 
parametrization proposed in Refs.~\cite{CGPmodel,CGP05}. As explained in 
detail in Secs.~\ref{systematics} and~\ref{parametrization}, these
two elements are intricately intertwined.

Our new framework comes with the price of introducing four new fit parameters
for each channel, the vector and axial-vector DV parameters. Nevertheless, we 
demonstrated that our method is feasible by presenting a rather complete 
analysis based on the OPAL data for the $V$ and $A$ spectral 
functions \cite{OPAL}. With the larger number of parameters, and the
corresponding need to vary $s_0$ away from $m_\t^2$,
it should come as no surprise that our errors are typically
larger than those found in earlier analyses, which 
simply did not take DVs into account quantitatively. 
We emphasize that this means that the systematic errors of 
those earlier analyses were understimated -- we believe significantly so in some
cases.

Our analysis leads us to a new estimate for both the central value of $\a_s$
and the error, which, in our opinion, should be interpreted as superseding
previous estimates in the literature; for our result, based on OPAL data,
see Sec.~\ref{summary}. We found our most reliable fits to be those of the $V$
channel, although fits including also the $A$ channel lead to results
consistent with our most precise $V$-channel fits. As our primary
concern in this article is with previously underestimated 
non-perturbative effects, we presented results for both 
contour-improved as well as fixed-order perturbation
theory. Our analysis was based on the original OPAL data, 
unmodified for subsequent changes in the various exclusive branching
fractions. This choice was made in order to facilitate interpretation
of the differences in our results from those obtained by OPAL. With this choice, 
these differences are solely the result of 
differences in the analysis method. We plan to present an analysis 
of OPAL data with updated normalizations in the near future.

The accuracy of the results presented here depends in part on our 
ability to correctly model the physics present in duality violations. 
Conservatively, our results can be
seen as providing a lower bound on the error introduced by ignoring 
duality violations. However, the stability of $\a_s$, a purely 
perturbative parameter, across the range of moments
analyzed here provides strong support for the validity of our \ansatz. 
We therefore surmise that not only is the \ansatz\ able to accurately 
describe the data, but also that it provides a reasonable quantitative 
description of the physics of duality violations in the
light-quark $V$ and $A$ channels.

In cases with multiple weights, standard $\c^2$ fits were not possible, and we performed 
alternate fits, propagating errors as described in App.~\ref{errorprop}. 
We point out, however, that our final result is based on 
a standard $\chi^2$ fit to the moment $I_{ex}^{(\hw_0)}$. 
All other fits yield results completely consistent with
this result, including the error on $\a_s$. 
The $\c^2$ error on $\a_s$ obtained from our fit to $I_{ex}^{(\hw_0)}$ 
is very close to that obtained with the method of App.~\ref{errorprop}.
Because our parameter covariance matrix obtained with that method 
scales linearly with the data covariance matrix, 
the error on $\a_s$ will be reduced once the improved spectral
function data expected from BaBar and/or Belle becomes available.

We observe that a difference remains between the central values for $\a_s$
obtained using CIPT and FOPT, though this is less significant than the 
difference found previously by OPAL, \seef\ Eq.~(\ref{OPALtau}). 
In this context, we note that, as mentioned already in Sec.~\ref{which},
a term linear in $s$ in any of the weights employed in Eq.~(\ref{cauchy})
picks out the $D=4$ term in the OPE,
which parametrizes the leading renormalon
ambiguity in the perturbative expansion.
This, then, raises the question whether differences 
between the behavior of CIPT and FOPT fits,
including those associated with any dependence on the choice of weight, 
might be used to constrain
renormalon pole models that have been used 
previously to investigate the resummation of the perturbative
series for the various sum rules employed in the study of hadronic $\t$ 
decays \cite{BJ,DM}.  We plan to pursue such an investigation in a future work.

\vspace{3ex}
\noindent {\bf Acknowledgments}
\vspace{3ex}

We would like to thank Martin Beneke, Claude Bernard, Andreas H\"ocker, 
Manel Martinez, and Ramon
Miquel for useful discussions.  We also would like to thank
Sven Menke for significant help with understanding the
OPAL spectral-function data.
MG thanks IFAE and the Department of Physics at UAB, and OC and KM thank the 
Department of Physics and Astronomy at SFSU for hospitality.
DB, MJ and SP are supported by CICYTFEDER-
FPA2008-01430, SGR2005-00916, the Spanish Consolider-Ingenio 2010 Program
CPAN (CSD2007-00042). SP is also supported by
 a fellowship from the Programa de Movilidad
PR2010-0284.
OC is supported in part by MICINN (Spain) under Grant FPA2007-60323, by the 
Spanish Consolider Ingenio 2010 Program CPAN (CSD2007-00042) and by the 
DFG cluster of excellence ``Origin and Structure of the Universe.''
MG and JO are supported in part by the US Department of Energy, and
KM is supported by a grant from the Natural Sciences and
Engineering Research Council of Canada.

\appendix
\section{\label{errorprop} Error propagation}
Consider a fit quality
\begin{equation}
\label{fitq}
{\cal Q}^2=\left[d_i-t_i({\vec p})\right]C^{-1}_{0,ij}\left[d_j-t_j({\vec p})\right]\ ,
\end{equation}
in which $d_i$ are the binned data, $t_i(\vec p)$ is a function that describes this
data set for a set of parameters $\vec p$, and $C_0$ is a positive-definite, symmetric,
but
otherwise arbitrary matrix. In this Appendix, we use the summation convention for
repeated indices.
 The parameters $\vec p$ are determined by finding
the global minimum of ${\cal Q}^2$, which satisfies
\begin{equation}
\label{minim}
\frac{\partial {\cal Q}^2}{\partial p_\a}=-2\;\frac{\partial t_i({\vec p})}{\partial p_\a}\;
C^{-1}_{0,ij}\left[d_j-t_j({\vec p})\right]=0\ .
\end{equation}
Varying, in this equation, the data by an amount $\d d_i$, and the parameters
by $\d p_\a$ leads to
\begin{equation}
\label{var}
\frac{\partial^2 t_i({\vec p})}{\partial p_\a\partial p_\b}\;C^{-1}_{0,ij}\left[d_j-t_j({\vec p})
\right]
\d p_\b+\frac{\partial t_i({\vec p})}{\partial p_\a}\;
C^{-1}_{0,ij}\left[\d d_j-\frac{\partial t_j({\vec p})}{\partial p_\b}\;\d p_\b\right]=0\ .
\end{equation}
If the fit is good, so that the deviations $d_j-t_j({\vec p})$ are small, we
may ignore the term with the second derivative, leading to
\begin{equation}
\label{deltap}
\d p_\a=A^{-1}_{\a\b}\frac{\partial t_i({\vec p})}{\partial p_\b}\;C^{-1}_{0,ij}
\;\d d_j\ ,
\end{equation}
or, for the covariance matrix $\langle\d p_\a\d p_\b\rangle$,
\begin{equation}
\label{parcorr}
\langle\d p_\a\d p_\b\rangle=A^{-1}_{\a\a'}A^{-1}_{\b\b'}
\frac{\partial t_i({\vec p})}{\partial p_{\a'}}
\frac{\partial t_j({\vec p})}{\partial p_{\b'}}
\;C^{-1}_{0,ik}C^{-1}_{0,j\ell}\;C_{k\ell}\ ,
\end{equation}
in which
\begin{equation}
\label{Adef}
A_{\a\b}=\frac{\partial t_i({\vec p})}{\partial p_\a}\;C^{-1}_{0,ij}\frac{\partial t_j({\vec p})}{\partial p_\b}\ ,
\end{equation}
and
\begin{equation}
\label{datacov}
C_{k\ell}=\langle\d d_k\d d_{\ell}\rangle
\end{equation}
is the data covariance matrix.
This provides us with an estimate for the full correlation matrix for the parameter set $\vec p$.  We note that, if $C_{0,ij}$ is chosen to be equal to the data covariance matrix $C_{ij}$, this expression simplifies to
\begin{equation}
\label{simple}
\langle\d p_\a\d p_\b\rangle=A^{-1}_{\a\b}\ .
\end{equation}
This is equal to the usual $\c^2$ error matrix estimate, given by the
inverse of one-half times the second derivative of ${\cal Q}^2$ at its minimum, if, again,
the fit is good enough to ignore terms proportional to $d_i-t_i({\vec p})$.


\end{document}